\documentclass[%reprint,
preprint,
amsmath,
amssymb,
aps,
]{revtex4-2}

\usepackage{graphicx}% Include figure files
\usepackage{dcolumn}% Align table columns on decimal point
\usepackage{bm}% bold math
\usepackage{amsfonts}
\usepackage{mathrsfs}
\usepackage{svg}
\begin{document}

\preprint{}

\title{Gravitational Faraday rotation of light propagation\\ in the Kerr–Newman–Taub–NUT space-time}% Force line breaks with

\author{Hongying Guo}
\email{21210180085@fudan.edu.cn}
\affiliation{School of Mathematical Sciences, Fudan University, Shanghai 200433, China}

\date{\today}% It is always \today, today,

\begin{abstract}
We investigate the gravitational Faraday effect in the Kerr-Newman-Taub-NUT space-time under the weak deflection limit. Contrary to previously stated zero net effect when the source and the observer are remote from the black hole, a non-zero Faraday rotation has been found. The rotation angle is dependent on the spin and the mass of the black hole and the observer's angular position, as in the case of the Kerr space-time, with additional contribution of the electrical charge and the NUT charge.
\end{abstract}

\maketitle \newpage

%\tableofcontents

\section{\label{Introduction}Introduction}

Light propagation in curved space-time can be influenced by the gravitational field in various ways. In addition to the gravitational lensing which changes the direction of light ray \cite{Cunha2018} and the gravitational red-shift of the photon \cite{Müller2010}, a mechanism has also been discovered to rotate the polarisation vector of the electromagnetic wave in gravitational field \cite{Ishihara1988,Nouri-Zonoz1999,Shoom2021,Kopeikin2002,Farooqui2014,Connors1977,Mashhoon1975,Volkov1970,Brodutch2011,Chakraborty2021,Sereno2004,Sereno2004DetectGravito}. This phenomenon is called the gravitational Faraday rotation, since it resembles the rotation of light polarisation in classical electromagnetics due to the magnetic field in medium.
To study the gravitational Faraday rotation in a curved space-time, the geometrical optics approximation is applied to the Maxwell equations under the Lorenz gauge, resulting in the polarisation vector parallelly transported along null geodesics and orthogonal to the wave vector \cite{Misner1973}. 

Although the rotation of polarisation in curved space-time has been noticed for a long time \cite{Volkov1970,Mashhoon1975,Plebanski1960}, 
it was not until the late 1970s when the gravitational Faraday effect of the Kerr black hole was first taken into consideration \cite{Connors1977,Stark1977}. 
The authors therein took advantage of the Petrov type-D property of the Kerr space-time, and utilised the existence of Walker-Penrose constant to determine the rotation of polarisation numerically. 
Later in \cite{Ishihara1988} the same technique was applied in the situation where both the source and the observer are sufficiently remote from the black hole, and calculated the Faraday rotation analytically.
Since then a number of authors paid attention to the Faraday rotation in Kerr space-time through various approaches \cite{Nouri-Zonoz1999,Tamburini2021,Farooqui2014,Brodutch2011,Gelles2021}. For instance, in \cite{Nouri-Zonoz1999} the stationary property of Kerr space-time renders it possible to project the null geodesic onto three-dimensional space, determining the angular velocity of Faraday rotation.
Both the results in \cite{Ishihara1988,Nouri-Zonoz1999} agree that the Faraday rotation in Kerr space-time should be proportional to the spin and the square of the mass of the black hole, in spite of the difference in their coefficients resulting from different definitions of Faraday rotation.
Besides the above mentioned results, an exact expression of the Faraday rotation in the Kerr geometry has been provided in \cite{Farooqui2014}.
Numerical studies have also been carried out recently for the determination of the Faraday rotation surrounding a Kerr black hole, such that the polarisation profile of the disk and jets around the black hole can be portrayed \cite{Chen2015,Dexter2016,Kawashima2021,Gelles2021}.

Increasing interest have also arisen towards the Faraday rotation in space-times other than the Kerr geometry \cite{Nouri-Zonoz1999,Chakraborty2021,Hou2019,Wang1991,Piran1985,Qin2022,Qin2022GB}. For the Taub-NUT space-time, the Faraday rotation angle turns out to be zero \cite{Nouri-Zonoz1999}. A non-zero Faraday rotation has been found in the magnetised Reissner-Nordström space-time \cite{Chakraborty2021}, which is a charged non-rotating black hole immersed in magnetic field, Ernst-transformed from standard Reissner-Nordström metric \cite{Ernst1976,Aliev1989}. For the magnetised Kerr space-time there exists a similar extra rotation due to the presence of magnetic field \cite{Chakraborty2021}. 
Besides, some authors have considered the Faraday rotation for space-times in modified theories of gravity \cite{Qin2022,Qin2022GB}.
Apart from these works on the Faraday rotation of light, the Faraday rotation of gravitational wave in curved space-times has also been studied \cite{Hou2019,Li2022,Wang1991,Piran1985}.

In this work we would like to investigate the gravitational Faraday rotation in the Kerr-Newman-Taub-NUT (KNTN) metric \cite{Griffiths2009}, which is a rotating black hole with spin $a$, electric charge $Q$ and NUT charge $l$. Our attention will be paid especially to the case with small spin, mass, electrical charge and NUT charge, and when the source and the observer are sufficiently remote. In a previous work \cite{Bini2003} a particular case with $Q = 0$ and $m$, $a$, $l$ being small has been considered, based on the method in \cite{Kopeikin2002}. However, their main concern was to derive the rotation angle for the source and the observer at a closer distance to the black hole, thus by keeping terms up to the first order of $a$ and $l$, they derived results inversely proportional to these distances. They concluded that through this method no Faraday rotation can be seen for sufficiently remote source and observer, and it is necessary to consider higher order effects for further investigation. Therefore, we prefer to take the full KNTN metric at the very beginning, and follow \cite{Ishihara1988} to use Taylor expansion up to the third order, so that the higher order Faraday rotation will not be omitted when the source and the observer are far away. We find a non-zero rotation angle dependent on the spin and the square of mass of the black hole, as in the case of Kerr space-time, with additional contribution of electrical charge and NUT charge.

The organisation of this paper is as follows. In Sec. \ref{Conserv KWP} we utilise the existence of Walker-Penrose constant in KNTN space-time to determine the propagation of polarisation vector. In Sec. \ref{derive faraday rotation} we choose the set of basis fitted for the realistic scenario of polarisation detection, and then calculate the Faraday rotation angle. The calculation is performed with the aid of the geodesic equations, under the weak deflection assumption. Discussions of the results are provided in Sec. \ref{discussions}. Details of the integration of geodesic equations are contained in the Appendix.

\section{\label{Conserv KWP} Formal derivation of the Faraday effect in KNTN space-time}\label{formal result}

In Boyer-Lindquist coordinates, the KNTN metric reads
\begin{equation}
\begin{aligned}
    \mathrm{d} s^{2} &=-\frac{1}{\Sigma }\left( \Delta -a^{2}\sin^{2} \theta \right)\mathrm{d} t^{2} \\
    &\ \ \ +\frac{2}{\Sigma }\left( \Delta p -a( \Sigma +a p )\sin^{2} \theta \right)\mathrm{d} t\mathrm{d} \phi\\
    &\ \ \ +\frac{1}{\Sigma }\left(( \Sigma +a p)^{2}\sin^{2} \theta -p ^{2} \Delta \right)\mathrm{d} \phi ^{2}\\
    &\ \ \ +\frac{\Sigma }{\Delta }\mathrm{d} r^{2} +\Sigma \mathrm{d} \theta ^{2},
\end{aligned}
\end{equation}
where
\begin{eqnarray*}
\begin{aligned}
    \Sigma &=r^{2} +( l+a\ \cos \theta )^{2},
    \\
    \Delta &=r^{2} -2mr-l^{2} +a^{2} +Q^{2},
    \\
    p &=a\sin^{2} \theta -2l\cos \theta.
\end{aligned}
\end{eqnarray*}
The parameters $m$, $a$, $Q$ and $l$ correspond to the mass, the spin, the electric charge and the NUT charge of the black hole, respectively.

Consider the geometrical optics approximation of the electromagnetic field in the KNTN space-time \cite{Misner1973}, where the polarisation vector $f$ and the wave vector $k$ should satisfy
\begin{eqnarray}\label{geometric approximation}
\begin{aligned}
    k \cdot k &= 0,\ 
    \nabla_{k}k = 0,\\
    k \cdot f &= 0,\ 
    \nabla_{k}f = 0.
\end{aligned}
\end{eqnarray}

The first two equations are the null condition and the geodesic equations governing the wave vector $k$, which can be treated through separation of variables \cite{Chandrasekhar1983}. For convenience of the readers the details of this procedure are included in the appendix. Then the complete form of the wave vector is
\begin{equation}\label{geodesic equation components}
\begin{aligned}
    k^{r} &=\pm \frac{\sqrt{R}}{\Sigma },\
    k^{\theta } =\pm \frac{\sqrt{\Theta }}{\Sigma },\\
    k^{t} &=\frac{1}{\Sigma \Delta }\left( r^{2} +a^{2} +l^{2}\right)\left( r^{2} +a^{2} +l^{2} -a\lambda \right) \\
    &\ \ \ +\frac{1}{\Sigma \sin^{2} \theta }\left( a\sin^{2} \theta -2l\cos \theta \right)\left( \lambda -a\sin^{2} \theta +2l\cos \theta \right),\\
    k^{\phi } &=\frac{1}{\Sigma \sin^{2} \theta }\left( \lambda -a\sin^{2} \theta +2l\cos \theta \right) +\frac{a}{\Sigma \Delta }\left( r^{2} +a^{2} +l^{2} -a\lambda \right),
\end{aligned}
\end{equation}
where
\begin{equation}
\begin{aligned}
    R&=-\Delta \left( \eta +( \lambda -a)^{2}\right) +\left( r^{2} +a^{2} +l^{2} -a\lambda \right)^{2}\\
    \Theta &=\eta +( \lambda -a)^{2} -\frac{1}{\sin^{2} \theta }\left( \lambda -a\sin^{2} \theta +2l\cos \theta \right)^{2},
\end{aligned}
\end{equation}
in which $\displaystyle\lambda = \frac{L_{z}}{E}$ is the reduced z-component of angular momentum, and $\displaystyle\eta = \frac{\mathcal{Q}}{E^2}$ is the reduced Carter constant of the null geodesic.

The last two equations of (\ref{geometric approximation}) are the orthogonality condition and the parallel transport equations of the polarisation vector $f$. 
Since the KNTN space-time is of Petrov type-D, after choosing a null tetrad-basis $(\mathfrak{l} ,\mathfrak{n} ,\mathfrak{m} ,\mathfrak{\overline{m}})$ with $\mathfrak{l}$ and $\mathfrak{n}$ being the principal null directions, the only non-vanishing component of Weyl curvature is $\Psi_{2}$. Therefore, there exists along the geodesic a conserved quantity called the Walker-Penrose constant \cite{Chandrasekhar1983}:
\begin{equation}\label{Walker Penrose definition}
\begin{aligned}
    K_{\mathrm{WP}} &=\{( k\cdot \mathfrak{l})( f\cdot \mathfrak{n}) -( k\cdot \mathfrak{n})( f\cdot \mathfrak{l}) -( k\cdot \mathfrak{m})( f\cdot \overline{\mathfrak{m}}) +( k\cdot \overline{\mathfrak{m}})( f\cdot \mathfrak{m})\} \Psi _{2}^{-1/3}\\
    &=2\{( k\cdot \mathfrak{l})( f\cdot \mathfrak{n}) -( k\cdot \mathfrak{m})( f\cdot \overline{\mathfrak{m}})\} \Psi _{2}^{-1/3}.
\end{aligned}
\end{equation}
Therefore, by inserting the wave vectors $k$ at any two locations along the same geodesic into (\ref{Walker Penrose definition}) respectively, the formal relation between polarisation vectors $f$ at these two locations can be automatically established, without solving the parallel transport equation.

Following \cite{Griffiths2009} we choose 
\begin{equation}\label{null tetrad}
\begin{aligned}
    \mathfrak{l} &=\frac{1}{\sqrt{2\Sigma }}\left(\frac{r^{2} +l^{2} +a^{2}}{\sqrt{\Delta }} \partial _{t} -\sqrt{\Delta } \partial _{r} +\frac{a}{\sqrt{\Delta }} \partial _{\phi }\right),\\
    \mathfrak{n} &=\frac{1}{\sqrt{2\Sigma }}\left(\frac{r^{2} +l^{2} +a^{2}}{\sqrt{\Delta }} \partial _{t} +\sqrt{\Delta } \partial _{r} +\frac{a}{\sqrt{\Delta }} \partial _{\phi }\right),\\
    \mathfrak{m} &=\frac{1}{\sqrt{2\Sigma }}\left(( a\sin \theta -2l\cot \theta ) \partial _{t} +\mathrm{i} \partial _{\theta } +\frac{1}{\sin \theta } \partial _{\phi }\right),\\
    \mathfrak{\overline{m}} &=\frac{1}{\sqrt{2\Sigma }}\left(( a\sin \theta -2l\cot \theta ) \partial _{t} -\mathrm{i} \partial _{\theta } +\frac{1}{\sin \theta } \partial _{\phi }\right),
\end{aligned}
\end{equation}
such that the expression of Walker-Penrose constant turns out to be
\begin{equation}\label{Walker Penrose constant value}
    K_{\textrm{WP}}=(A+\mathrm{i}B)\Psi_{2}^{-1/3},
\end{equation}
with
\begin{eqnarray*}
\begin{aligned}
    A &= k^{t}f^{r}-k^{r}f^{t}
    +(a\sin^2\theta-2l\cos\theta)(k^{r}f^{\phi}-k^{\phi}f^{r}),\\
    B &= (r^2+a^2+l^2)\sin\theta(k^{\phi}f^{\theta}-k^{\theta}f^{\phi})-a\sin\theta(k^{t}f^{\theta}-k^{\theta}f^{t}),\\
    \Psi_{2} &= \left(r-\mathrm{i}(l+a\cos\theta)\right)^{-3}
    \left(-m+\mathrm{i}l+\frac{Q^2}{r+\mathrm{i}(l+a\cos\theta)}\right).
\end{aligned}
\end{eqnarray*}

Further assume that the source and observer are sufficiently far away from the black hole, then the value of the wave vectors  $k_{\textrm{s}}$ and $k_{\textrm{o}}$ can be extracted from the asymptotics of the geodesic equations (\ref{geodesic equation components}):
\begin{eqnarray}\label{wave vector asymptotics}
\begin{aligned}
k^{t} &=1+O\left(\frac{1}{r}\right),\ 
k^{r} =\pm 1+O\left(\frac{1}{r}\right),\\
k^{\theta } &=\frac{\beta }{r^{2}} +O\left(\frac{1}{r^{3}}\right),\ 
k^{\phi } =\frac{\lambda +2l\cos \theta }{r^{2}\sin^{2} \theta } +O\left(\frac{1}{r^{3}}\right),
\end{aligned}
\end{eqnarray}
where
\begin{equation}\label{beta and gamma}
\begin{aligned}
    \beta &= \pm \left(\eta + (\lambda-a)^2 - (2l\cot\theta + \lambda\csc\theta - a\sin\theta)^2\right)^{1/2},\\
    \gamma &= \lambda\csc\theta - a\sin\theta.
\end{aligned}
\end{equation}

With these asymptotics in mind, the Walker-Penrose constant for the sufficiently remote source and observer leads to the relation
\begin{equation}\label{Walker Penrose conservation}
    (\beta \hat{f}^{\theta } +\gamma \hat{f}^{\phi })
    _{\textrm{s}}
    +\mathrm{i}(\gamma \hat{f}^{\theta } -\beta \hat{f}^{\phi })
    _{\textrm{s}}
    =-(\beta\hat{f}^{\theta } +\gamma\hat{f}^{\phi })_{\textrm{o}}
    +\mathrm{i}(\gamma\hat{f}^{\theta } -\beta \hat{f}^{\phi })_{\textrm{o}},
\end{equation}
where we have also utilised the gauge condition $f^{t} \equiv 0$, the orthogonality condition $k \cdot f = 0$, and the notations
\begin{eqnarray}\label{notation of polarisation}
\begin{aligned}
    \hat{f}^{r} &= f^{r},
    \hat{f}^{\theta } = r f^{\theta},
    \hat{f}^{\phi } = r\sin\theta f^{\phi}. 
\end{aligned}
\end{eqnarray}
This relation regarding $\hat{f}^{\theta }$ and $\hat{f}^{\phi }$ reveals the complete transformation of polarisation vector, since the asymptotics of $k$ combined with the orthogonality condition indicate that $\hat{f}^{r}$ vanishes for remote source and observer.

Note that (\ref{Walker Penrose conservation}) is complex, then by solving $\hat{f}^{\theta }_{\textrm{o}}$ and $\hat{f}^{\phi }_{\textrm{o}}$ from the real and imaginary parts, we have
\begin{equation}\label{transformation of f theta f phi}
\begin{pmatrix}
\hat{f}^{\theta }\\
\hat{f}^{\phi }
\end{pmatrix}_{\mathrm{o}} 
=\mathcal{R} \begin{pmatrix}
\hat{f}^{\theta }\\
\hat{f}^{\phi }
\end{pmatrix}_{\mathrm{s}},
\end{equation}
where
\begin{equation}\label{original transformation matrix}
\begin{aligned}
\mathcal{R} 
&=\frac{1}{\sqrt{1+x^{2}}}\begin{pmatrix}
1 & -x\\
-x & -1
\end{pmatrix},\\
x&=\frac{\beta _{\mathrm{s}} \gamma _{\mathrm{o}} +\gamma _{\mathrm{s}} \beta _{\mathrm{o}}}{\gamma _{\mathrm{s}} \gamma _{\mathrm{o}} -\beta _{\mathrm{s}} \beta _{\mathrm{o}}}.
\end{aligned}
\end{equation}

Equation (\ref{transformation of f theta f phi}) describes the transformation from the polarisation vector at the source to that at the observer. However, the result presented here is based on the global coordinates $\theta$ and $\phi$ of the space-time manifold, and it is difficult to experimentally compare the respective components $\hat{f}^\theta$ and $\hat{f}^\phi$ at the source and the observer.
Only after choosing meaningful polarisation frames at these two locations, can we compare the polarisation vectors and talk about the rotation of polarisation during the course of deflection. 

\section{Derivation of the gravitational Faraday rotation angle}\label{derive faraday rotation}

In section \ref{formal result}, we have derived the formal transformation matrix $\mathcal{R}$ of the polarisation vector corresponding to the global coordinate system.
In this section, we will carry out detailed calculations of the polarisation rotation angle after choosing the observationally favourable bases for the source and the observer respectively.
We have assumed that both the source and the observer are located in the Minkowskian far zone, in which the trajectory of light can be seen as lying in an orbital plane. Consequently, the observer can experimentally detect the direction of light and determine the normal $\mathbf{n}$ of this orbital plane. 
We set $\mathbf{n}$ as one component of the polarisation basis at the observer, then the other component $\mathbf{h}_{\mathrm{o}}$ is fixed by requiring that $\mathbf{h}_{\mathrm{o}}$ being orthogonal both to $\mathbf{n}$ and to the wave vector $\mathbf{k}_{\mathrm{o}}$.
For the source the same vector $\mathbf{n}$ is adopted in the polarisation basis, accompanied by the other component $\mathbf{h}_{\mathrm{s}}$ which is orthogonal to $\mathbf{n}$ and $\mathbf{k}_{\mathrm{s}}$.

\begin{figure}[h]
    \centering
    \includegraphics[width=0.7\textwidth]{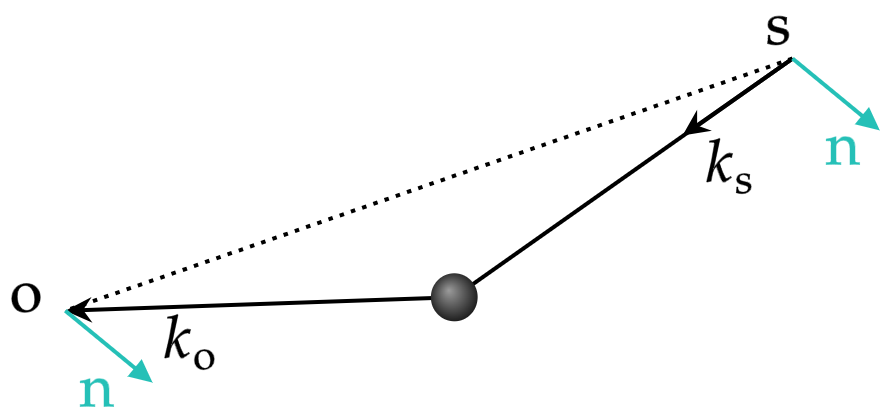}
    \caption{Demonstration of the orbital plane of light and its normal $\mathbf{n}$. The wave vectors at the source and the observer are depicted. Details of light ray in the region strongly influenced by gravitation is neglected, which is represented by the dark area in the centre.}
    \label{fig1}
\end{figure}

Following \cite{Ishihara1988} we write in Euclidean coordinates
\begin{eqnarray}\label{definition new frame}
\begin{aligned}
    \mathbf{n} &=\mathbf{k}_{\mathrm{o}} \times 
    \mathbf{k}_{\mathrm{s}},\\
    \mathbf{h}_{\mathrm{s}} &=\mathbf{n}\times \mathbf{k}_{\mathrm{s}},\\
    \mathbf{h}_{\mathrm{o}} &=\mathbf{n}\times \mathbf{k}_{\mathrm{o}}
\end{aligned}
\end{eqnarray}
where
\begin{eqnarray}
\begin{aligned}
    \mathbf{k}_{\mathrm{o}} &=(\sin \theta _{\mathrm{o}}\cos \phi _{\mathrm{o}} ,\sin \theta _{\mathrm{o}}\sin \phi _{\mathrm{o}} ,\cos \theta _{\mathrm{o}}),\\
    \mathbf{k}_{\mathrm{s}} &=-(\sin \theta _{\mathrm{s}}\cos \phi _{\mathrm{s}} ,\sin \theta _{\mathrm{s}}\sin \phi _{\mathrm{s}} ,\cos \theta _{\mathrm{s}}).
\end{aligned}
\end{eqnarray}

\begin{figure}[h]
    \centering
    \includegraphics[width=0.7\textwidth]{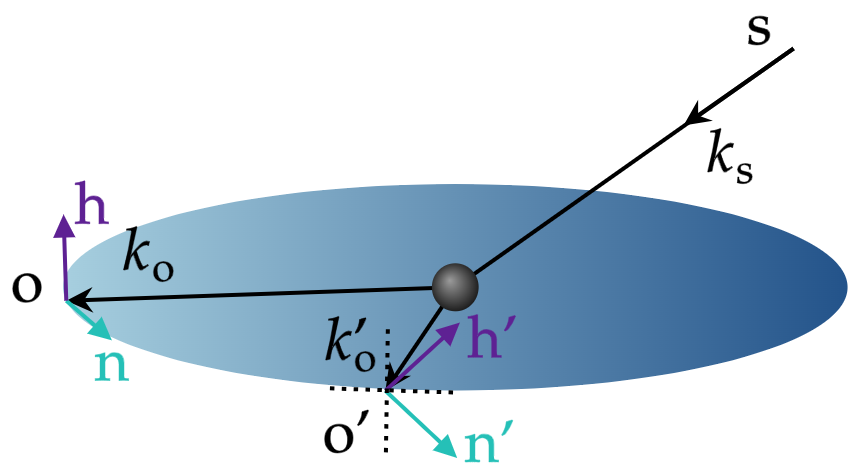}
    \caption{Demonstration of different polarisation bases for light emitted from the same source and reaching observer o and observer o'. The straight lines with arrows depict light propagation in the Minkowskian far zone. Details of light ray in the region influenced by gravitation is neglected, represented by the dark area in the centre. Note the differences between the bases at two observers.}
    \label{fig1}
\end{figure}

Since 
\begin{equation}
    \mathbf{f}
    =f^{\parallel}\mathbf{h}+f^{\perp}\mathbf{n}
    =\hat{f}^{\theta} \mathbf{e}_{\theta} + \hat{f}^{\phi} \mathbf{e}_{\phi},
\end{equation}
where $\mathbf{e}_{\theta }$ and $\mathbf{e}_{\phi }$, are the unit vectors along the directions of $\theta$ and $\phi$ in Euclidean space, it follows that
\begin{equation}\label{new transformation of linear light}
\begin{aligned}
\begin{pmatrix}
f_{\parallel }\\
f_{\perp }
\end{pmatrix}_{\mathrm{o}} 
&=N_{\mathrm{o}} \mathcal{R} N_{\mathrm{s}}^{-1}
\begin{pmatrix}
f_{\parallel }\\
f_{\perp }
\end{pmatrix}_{\mathrm{s}},
\end{aligned}
\end{equation}
where the transformation matrices $N_{\mathrm{o}}$ and $N_{\mathrm{s}}$ are
\begin{eqnarray}\label{transformation matrices Ns No}
\begin{aligned}
N_{\mathrm{s}} =\begin{pmatrix}
h^{\theta } & h^{\phi }\\
n^{\theta } & n^{\phi }
\end{pmatrix}_{\mathrm{s}},\ 
N_{\mathrm{o}} =\begin{pmatrix}
h^{\theta } & h^{\phi }\\
n^{\theta } & n^{\phi }
\end{pmatrix}_{\mathrm{o}},    
\end{aligned}
\end{eqnarray}
where we have
\begin{equation}\label{relationship h and n}
    h_{\mathrm{s}}^{\theta } = -n_{\mathrm{s}}^{\phi },\ 
    h_{\mathrm{s}}^{\phi } = n_{\mathrm{s}}^{\theta },\
    h_{\mathrm{o}}^{\theta } = n_{\mathrm{s}}^{\phi },\
    h_{\mathrm{o}}^{\phi } = -n_{\mathrm{s}}^{\theta },
\end{equation}
and the components of $\mathbf{n}$ with respect to $\mathbf{e}_{\theta }$ and $\mathbf{e}_{\phi }$ are
\begin{eqnarray}\label{components of n}
\begin{aligned}
    n_{\mathrm{s}}^{\theta } &=\sin \theta _{\mathrm{o}}\sin( \phi _{\mathrm{s}} -\phi _{\mathrm{o}}),\\
    n_{\mathrm{s}}^{\phi } &=\sin \theta _{\mathrm{o}}\cos \theta _{\mathrm{s}}\cos( \phi _{\mathrm{s}} -\phi _{\mathrm{o}}) -\sin \theta _{\mathrm{s}}\cos \theta _{\mathrm{o}},\\
    n_{\mathrm{o}}^{\theta } &=\sin \theta _{\mathrm{s}}\sin( \phi _{\mathrm{s}} -\phi _{\mathrm{o}}),\\
    n_{\mathrm{o}}^{\phi } &=-\sin \theta _{\mathrm{s}}\cos \theta _{\mathrm{o}}\cos( \phi _{\mathrm{s}} -\phi _{\mathrm{o}}) +\sin \theta _{\mathrm{o}}\cos \theta _{\mathrm{s}}.
\end{aligned}
\end{eqnarray}

After taking account of the equations (\ref{original transformation matrix})(\ref{new transformation of linear light})(\ref{transformation matrices Ns No})(\ref{relationship h and n}), we realise that the final transformation matrix $N_{\mathrm{o}} \mathcal{R} N_{\mathrm{s}}^{-1}$ is a rotation matrix, such that
\begin{equation}\label{new transformation of linear light 2}
\begin{aligned}
\begin{pmatrix}
f_{\parallel }\\
f_{\perp }
\end{pmatrix}_{\mathrm{o}}
&=\begin{pmatrix}
\cos\chi & -\sin\chi\\
\sin\chi & \cos\chi
\end{pmatrix}
\begin{pmatrix}
f_{\parallel }\\
f_{\perp }
\end{pmatrix}_{\mathrm{s}},    
\end{aligned}
\end{equation}
where $\chi$ is the Faraday rotation angle defined by
\begin{equation}\label{definition of chi}
\begin{aligned}
    \sin\chi &= \frac{X-x}{\sqrt{(1+X^2)(1+x^2)}},\\
    X &=\frac{n_{\mathrm{o}}^{\theta}n_{\mathrm{s}}^{\phi}+n_{\mathrm{o}}^{\phi}n_{\mathrm{s}}^{\theta}}{n_{\mathrm{o}}^{\phi}n_{\mathrm{s}}^{\phi}-n_{\mathrm{o}}^{\theta}n_{\mathrm{s}}^{\theta}}.
\end{aligned}
\end{equation}
Here $x$ has been defined in (\ref{original transformation matrix}), and the values of $n_{\mathrm{s}}^{\theta } $, $n_{\mathrm{s}}^{\phi } $, $n_{\mathrm{o}}^{\theta } $, $n_{\mathrm{o}}^{\phi } $ have been provided in (\ref{components of n}).

Since it is difficult to analytically derive the value of the Faraday rotation angle for the general cases, now we will focus on the simplest and physically most relevant case, where it is assumed that the whole light ray propagates in the far zone of the black hole, i.e., when
\begin{equation}\label{weak deflection}
    \frac{\mathrm{max}\{m,\vert a \vert, \vert Q \vert, \vert l \vert\}}{r_{\mathrm{min}}} \ll 1,
\end{equation}
such that there is only weak deflection of the light during propagation
\begin{eqnarray}
\begin{aligned}
\Delta\theta = \theta_{\mathrm{o}} + \theta_{\mathrm{s}} - \pi &\ll 1,\\
\Delta\phi = \phi_{\mathrm{o}} - \phi_{\mathrm{s}} - \pi &\ll 1.
\end{aligned}
\end{eqnarray}
Besides, since the source and observer are both remote, the terms of order $r_{\mathrm{min}}/\mathrm{min}\{r_{\mathrm{o}},r_{\mathrm{s}}\}$ are supposed to be small quantities of even higher order and will be ignored.
We also require that the null geodesic reaches the maximum or minimum of $\theta$ only once for simplicity, with the existence of $\theta_{\mathrm{max}}$ or $\theta_{\mathrm{min}}$ provided by the solution (\ref{cos theta min}) of the equation $\Theta(\theta)=0$ when $\eta \geq 0$ \cite{Chandrasekhar1983}. Note that when $\theta_{\mathrm{min}}$ is reached, the wave vector at the source must satisfy $k^{\theta}_{\mathrm{s}}<0$, thus $\beta_{\mathrm{s}}<0$ due to (\ref{wave vector asymptotics}), while at the observer we have $k^{\theta}_{\mathrm{o}}>0$ and $\beta_{\mathrm{o}}>0$. If $\theta_{\mathrm{max}}$ is reached instead, then the converse sign should be taken.

\begin{figure}[h]
    \centering
    \includegraphics[width=0.6\textwidth]{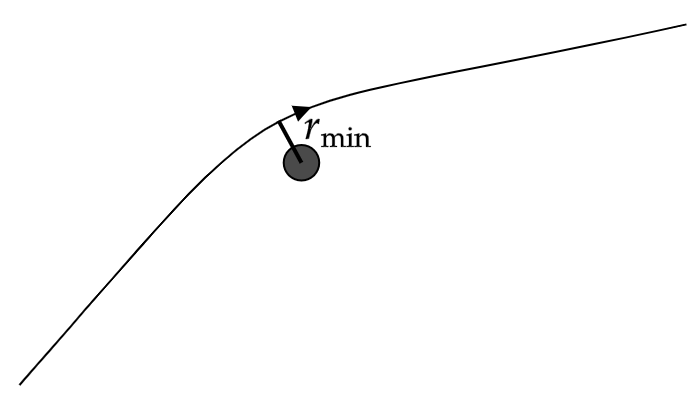}
    \caption{The minimal value of the radial coordinate along trajectory $r_{\mathrm{min}}$ should be significantly larger than the parameters of the black hole. Deflection angle has been exaggerated. The dark region represents the black hole.}
    \label{fig2}
\end{figure}

Under such assumptions we immediately see that $X$ can be calculated approximately by first inserting 
\begin{equation}\label{weak deflection}
\begin{aligned}
    \theta_{\mathrm{s}} = \pi - \theta_{\mathrm{o}} + \Delta\theta,\\
    \phi_{\mathrm{s}}  = \phi_{\mathrm{o}} - \pi -\Delta\phi
\end{aligned}
\end{equation}
into the second equation in (\ref{definition of chi}) with considerations of the expressions (\ref{components of n}), and then expanding the result with respect to $\Delta\theta$ and $\Delta\phi$ up to the third order. In fact, this has already been done in \cite{Ishihara1988}, and the same result can be derived here:
\begin{equation}\label{expansion X}
\begin{aligned}
    X  
    = \cos\theta_{\mathrm{o}} \Delta\phi 
    +\frac{1}{2}\sin\theta_{\mathrm{o}} \Delta\theta\Delta\phi
    +\frac{1}{12}(1+3\cos^2\theta_{\mathrm{o}})\cos\theta_{\mathrm{o}} \Delta\phi^3.
\end{aligned}
\end{equation}

For the expression of $x$, taking the similar approach, we have from equations (\ref{beta and gamma}) and (\ref{weak deflection})
\begin{equation}
\begin{aligned}
    \beta _{\mathrm{s}}
    &=\mp \left(\eta + (\lambda-a)^2 - (2l\cot\theta_{\mathrm{s}} + \lambda\csc\theta_{\mathrm{s}} - a\sin\theta_{\mathrm{s}})^2\right)^{1/2}\\
    &=\mp \left(\eta + (\lambda-a)^2 
    - (-2l\cot(\theta_{\mathrm{o}}-\Delta\theta) + \lambda\csc(\theta_{\mathrm{s}}-\Delta\theta) - a\sin(\theta_{\mathrm{s}}-\Delta\theta))^2\right)^{1/2},\\
    \beta _{\mathrm{o}}
    &=\pm \left(\eta + (\lambda-a)^2 - (2l\cot\theta_{\mathrm{o}} + \lambda\csc\theta_{\mathrm{o}} - a\sin\theta_{\mathrm{o}})^2\right)^{1/2},\\
    \gamma_{\mathrm{s}}
    &=\lambda\csc\theta_{\mathrm{s}} - a\sin\theta_{\mathrm{s}}
    =\lambda\csc(\theta_{\mathrm{o}}-\Delta\theta) - a\sin(\theta_{\mathrm{o}}-\Delta\theta),\\
    \gamma_{\mathrm{o}}
    &=\lambda\csc\theta_{\mathrm{o}} - a\sin\theta_{\mathrm{o}},
\end{aligned}
\end{equation}
such that
\begin{equation}\label{expansion x}
\begin{aligned}
    x & 
    = \frac{\beta _{\mathrm{s}} \gamma _{\mathrm{o}} +\gamma _{\mathrm{s}} \beta _{\mathrm{o}}}{\gamma _{\mathrm{s}} \gamma _{\mathrm{o}} -\beta _{\mathrm{s}} \beta _{\mathrm{o}}}\\
    & = \pm \Big\{
    -4\tilde{l}\cos \theta _{\mathrm{o}} \mu ^{-1} +2\tilde{a}^{2}\tilde{l}\sin^{2} \theta _{\mathrm{o}}\cos^{3} \theta _{\mathrm{o}} \mu ^{-3} +4\tilde{l}^{3}\cos^{3} \theta _{\mathrm{o}}( 3\cos 2\theta _{\mathrm{o}} -3\tilde{\eta } +\tilde{\lambda }^{2}) \mu ^{-5}\\
    &\ \ \ +\Delta\theta\Big(
    \tilde{\lambda }\cot \theta _{\mathrm{o}} \mu ^{-1} +\tilde{a}\sin \theta _{\mathrm{o}}\cos \theta _{\mathrm{o}} \mu ^{-1} -2\tilde{\eta }\tilde{l}\sin \theta _{\mathrm{o}} \mu ^{-3} -\frac{1}{2}\tilde{a}^{2}\tilde{\lambda }\sin^{4} \theta _{\mathrm{o}}\cot^{3} \theta _{\mathrm{o}} \mu ^{-3}\\
    &\ \ \ -\frac{1}{4}\tilde{\lambda }\tilde{l}^{2}\cot \theta _{\mathrm{o}}\big( 8(\tilde{\eta } +4\tilde{\lambda }^{2})\cos 2\theta _{\mathrm{o}} +9\cos 4\theta _{\mathrm{o}} -17\tilde{\eta } +23\tilde{\lambda }^{2}\big) \mu ^{-5}\Big)\\
    &\ \ \ +\Delta\theta^{2}\Big(
    -\frac{1}{2}\tilde{\lambda }\sin^{2} \theta _{\mathrm{o}}( -2\csc^{2} \theta _{\mathrm{o}} +1 +\tilde{\lambda }^{2}\csc^{4} \theta _{\mathrm{o}} ) \mu ^{-3}\\
    &\ \ \ +\frac{1}{2}\tilde{a}\sin^{4} \theta _{\mathrm{o}}(1 +\tilde{\lambda }^{2}\cos 2\theta _{\mathrm{o}}\csc^{4} \theta _{\mathrm{o}} ) \mu ^{-3}\\
    &\ \ \ +\frac{1}{4}\tilde{l}\cot \theta _{\mathrm{o}}\csc \theta _{\mathrm{o}}\big(( 4\tilde{\eta }^{2} +6\tilde{\eta }\tilde{\lambda }^{2} +8\tilde{\lambda }^{4})\cos 2\theta _{\mathrm{o}} \\
    &\ \ \ -3\tilde{\eta }^{2} -(\tilde{\eta } -2\tilde{\lambda }^{2})\cos 4\theta _{\mathrm{o}} -7\tilde{\eta }\tilde{\lambda }^{2} +6\tilde{\lambda }^{4}\big) \mu ^{-5}\Big)\\
    &\ \ \ +\Delta\theta^{3}\Big(
    -\frac{1}{48}\tilde{\lambda }\cot \theta _{\mathrm{o}}\big( 4( 5\tilde{\eta }^{2} +6\tilde{\eta }\tilde{\lambda }^{2} +\tilde{\lambda }^{4})\cos 2\theta _{\mathrm{o}} \\
    &\ \ \ -21\tilde{\eta }^{2} +\cos 4\theta _{\mathrm{o}} -2\tilde{\eta }\tilde{\lambda }^{2} +3\tilde{\lambda }^{4}\big) \mu ^{-5}
    \Big)\Big\},
\end{aligned}
\end{equation}
where we defined
\begin{eqnarray}\label{tilde notations}
\begin{aligned}
    \tilde l &= \frac{l}{\sqrt{\lambda^2 + \eta}},\ 
    \tilde a = \frac{a}{\sqrt{\lambda^2 + \eta}},\
    \tilde \lambda &= \frac{\lambda}{\sqrt{\lambda^2 + \eta}},\ 
    \tilde \eta = \frac{\eta}{\lambda^2 + \eta},\
    \mu &= \sqrt{\tilde\eta - \cos^2\theta_{\mathrm{o}}},
\end{aligned}
\end{eqnarray}
and kept the third order terms of $\tilde a$, $\tilde l$, and $\Delta\theta$ in our calculation.
It turns out that both $X$ and $x$ are small quantities, therefore, the Faraday rotation angle $\chi$ satisfy
\begin{equation}\label{chi approx sine of chi}
\chi \approx \sin\chi 
    = \frac{X-x}{\sqrt{(1+X^2)(1+x^2)}}.
\end{equation}
In fact, we will find out later that after plugging in the expressions of $\Delta\theta$ and $\Delta\phi$ for the chosen geodesics, $X$ and $x$ cancel each other up to the second order of $\tilde m$, $\tilde a$, $\tilde Q$ and $\tilde l$, such that $X-x$ is a small quantity of the third order. If we are only interested in the third order term of Faraday rotation with respect to these small quantities, the rotation angle can be calculated by
\begin{equation}
    \chi = X - x .
\end{equation}

Among all the weakly deflected geodesics, we only consider here the geodesics either reaching $\theta_{\mathrm{min}}$ or $\theta_{\mathrm{max}}$ only once for simplicity.
We will see that since the integral form of geodesic equations establishes the link between the values of $\theta$ and $\phi$ at the source and the observer, it is feasible to express $\Delta\theta$ and $\Delta\phi$ as functions of $\theta_{\mathrm{o}}$, $\lambda$, $\eta$, and the small quantities $\tilde m$, $\tilde a$, $\tilde Q$ and $\tilde l$.
Therefore, when inserting these expressions back into $X$ and $x$ in (\ref{expansion X}) and (\ref{expansion x}), the results will no longer explicitly depend on $\theta_{\mathrm{s}}$, $\phi_{\mathrm{s}}$ and $\theta_{\mathrm{o}}$, and will be much simpler.

By eliminating the affine parameter in (\ref{r theta geodesic}) and (\ref{conserve E Lz}), the integral form of the spatial parts of the geodesic equations are
\begin{equation}\label{spatial integration}
\begin{aligned}
    &\int ^{r}\frac{\mathrm{| d r| }}{\sqrt{R( r)}} =\int ^{\theta }\frac{\mathrm{| d\theta | }}{\sqrt{\Theta ( \theta )}},\\
    &\phi _{\mathrm{o}} -\phi _{\mathrm{s}} 
    =\int ^{r}\frac{a\left( r^2+a^2+l^2-a\lambda\right)\mathrm{| d r| }}{\Delta \sqrt{R( r)}}
    +\int ^{\theta }\frac{\left( \lambda \csc^{2} \theta -a +2l\csc^{2} \theta \cos \theta \right)\mathrm{| d\theta | }}{\sqrt{\Theta ( \theta )}}\\
    &\ \ \ \ \ \ \ \ \ \ =\int ^{r}\frac{a\left( 2mr-a \lambda -Q^{2} +2l^{2}\right)\mathrm{| d r| }}{\Delta \sqrt{R( r)}}
    +\int ^{\theta }\frac{\left( \lambda \csc^{2} \theta +2l\csc^{2} \theta \cos \theta \right)\mathrm{| d\theta | }}{\sqrt{\Theta ( \theta )}},
\end{aligned}
\end{equation}
where
\begin{equation}
\begin{aligned}
    R( r) &=-\Delta \left( \eta +( \lambda -a)^{2}\right)
    +\left( a\lambda -\left( r^{2} +l^{2} +a^{2}\right)\right)^{2},\\
    \Theta ( \theta ) &=\eta +( \lambda -a)^{2}
    -\frac{1}{\sin^{2} \theta }\left( \lambda -a\sin^{2} \theta +2l\cos \theta \right)^{2}.
\end{aligned}
\end{equation}
The integrals are taken from the source along the geodesic to the observer, with the requirement that $\theta$ reaches its minimum or maximum only once,
while the range of $r$ is from $r_{\mathrm{s}}$ to $r_{\mathrm{min}}$ then back to $r_{\mathrm{o}}$.
Note that the second equation in (\ref{spatial integration}) has been simplified by the first equation therein. 

The integration involving $r$ will be directly calculated after the substitution of variable and Taylor expansion with respect to the small quantities. On the other hand, the treatment of integration involving $\theta$ will be much more complicated. Since the range of the integration is determined by $\theta_{\mathrm{s}}$ and $\theta_{\mathrm{o}}$, which are related by $\Delta\theta$ as in (\ref{weak deflection}), the first equation in (\ref{spatial integration}) will eventually become a polynomial equation over $\Delta\theta$, from which the value of $\Delta\theta$ can be solved. Then by inserting the expression of $\Delta\theta$ into the second equation of (\ref{spatial integration}), the value of $\Delta\phi$ can also be determined.

Now we proceed to calculate the first integration with respect to $r$. Since the source and the observer are sufficiently remote from the black hole, or equivalently, both $r_{\mathrm{s}}$ and $r_{\mathrm{o}}$ are large enough, we have
\begin{equation}
\begin{aligned}
    \int ^{r}\frac{\mathrm{| d} r\mathrm{| }}{\sqrt{R( r)}} 
    &=2\int _{r_{\mathrm{min}}}^{\infty }\frac{\mathrm{d} r}{\sqrt{R( r)}}
    -\int _{r_{\mathrm{s}}}^{\infty}\frac{\mathrm{d} r}{\sqrt{R( r)}}
    -\int _{r_{\mathrm{o}}}^{\infty}\frac{\mathrm{d} r}{\sqrt{R( r)}}\\
    & \approx 2\int _{r_{\mathrm{min}}}^{\infty }\frac{\mathrm{d} r}{\sqrt{R( r)}}
    -\frac{1}{r_{\mathrm{s}}}
    -\frac{1}{r_{\mathrm{o}}}\\
    & \approx 2\int _{r_{\mathrm{min}}}^{\infty }\frac{\mathrm{d} r}{\sqrt{R( r)}}.
\end{aligned}
\end{equation}
Here the value of $r_{\mathrm{min}}$ should be determined before evaluating the integral. From $R(r_{\mathrm{min}})=0$ we have up to the second order of the small quantities $\tilde m$, $\tilde a$, $\tilde Q$ and $\tilde l$
\begin{equation}
    r_{\mathrm{min}} =r_{\mathrm{min}}^{( 0)}\Bigg( 1-\tilde{m} -\frac{1}{2}\tilde{\lambda }\tilde{a}^{2} -\frac{3}{2}\tilde{m}^{2} +\frac{1}{2}\tilde{Q}^{2} 
    -\frac{3}{2}\tilde{l}^{2} +2\tilde{\lambda }\tilde{a}\tilde{m}\Bigg),
\end{equation}
with 
\begin{eqnarray*}
    r_{\mathrm{min}}^{( 0)}=\sqrt{\lambda^2+\eta},\ 
    \tilde m = \frac{m}{\sqrt{\lambda^2 + \eta}},\ 
    \tilde Q = \frac{Q}{\sqrt{\lambda^2 + \eta}}
\end{eqnarray*}
besides the notations in (\ref{tilde notations}).

By the substitution of variable as (\ref{substitution of new variable}) and (\ref{substitution of new variable integration}) in the appendix, and Taylor expansion of the integrand up to the third order of $\tilde a$, $\tilde m$, $\tilde Q$ and $\tilde l$, the result of the integration is
\begin{equation}\label{spatial integration r1}
\begin{aligned}
    \int ^{r}\frac{\mathrm{| d} r\mathrm{| }}{\sqrt{R( r)}} 
    &= r{_{\mathrm{min}}^{( 0)}}^{-1}
    \Big(
     \pi+4\tilde{m}+32\tilde{m}\tilde{l}^{2} -16\tilde{m}\tilde{Q}^{2} -8\tilde{m}\tilde{a}\tilde{\lambda } 
    +4\tilde{m}\tilde{a}^{2}( 4\tilde{\lambda }^{2} -1)\\
    &\ \ \ +\frac{7}{4}\pi \tilde{l}^{2} 
    -\frac{3}{4} \pi \tilde{Q}^{2}
    -3\pi\tilde{a}\tilde{l}^{2}\tilde{\lambda }
    +\frac{3}{2}\pi\tilde{a}\tilde{Q}^{2}\tilde{\lambda }
    -\frac{1}{4}\pi\tilde{a}^{2}
    +\frac{3}{4}\pi\tilde{\lambda }^{2}\tilde{a}^{2}\\
    &\ \ \ +\frac{128}{3}\tilde{m}^{3} +\frac{15}{4} \pi \tilde{m}^{2} -15\pi \tilde{m}^{2}\tilde{a}\tilde{\lambda }\Big).
\end{aligned}    
\end{equation}
For the integration involving $r$ in the second equation of (\ref{spatial integration}), the result is
\begin{equation}
\begin{aligned}
    &\ \ \ \int ^{r}\frac{\left( 2mar-a^{2} \lambda -aQ^{2} +2al^{2}\right)\mathrm{| d r| }}{\Delta \sqrt{R( r)}} \\
    &= 4\tilde{a}\tilde{m} +5\pi \tilde{a}\tilde{m}^{2} -\frac{\pi }{2}\tilde{a}^{2}\tilde{\lambda } -8\tilde{a}^{2}\tilde{m}\tilde{\lambda } -\frac{\pi }{2}\tilde{a}\tilde{Q}^{2} +\pi \tilde{a}\tilde{l}^{2}.
\end{aligned}  
\end{equation}

For the integration over $\theta$, the range of integral is from $\theta_{\mathrm{s}}$ to $\theta_{\mathrm{min}}$ or $\theta_{\mathrm{max}}$, and then to $\theta_{\mathrm{o}}$. In the following calculations we will only take the geodesics reaching $\theta_{\mathrm{min}}$ for illustration. 

The value of $\theta_{\mathrm{min}}$ is first derived from $\Theta(\theta_{\mathrm{min}})=0$, up to the third order of $\tilde a$ and $\tilde l$:
\begin{equation}\label{cos theta min}
\begin{aligned}
    \cos \theta _{\mathrm{min}} &=\tilde{\eta }^{1/2}
    \Big( 1-2\tilde{\lambda }\tilde{\eta }^{-1/2}\tilde{l} +\frac{1}{2}\tilde{\lambda }^{2}\tilde{a}^{2}-2( 1-\tilde{\lambda }^{2}\tilde{\eta }^{-1})\tilde{l}^{2} +2\tilde{\lambda }^{2}\tilde{\eta }^{-1/2}\tilde{a}\tilde{l}\\
    &\ \ \ +2( 2\tilde{\lambda }\tilde{\eta }^{1/2} -\tilde{\lambda }\tilde{\eta }^{-1/2})\tilde{a}^{2}\tilde{l}
    +4( 3\tilde{\lambda } -\tilde{\lambda }\tilde{\eta }^{-1})\tilde{a}\tilde{l}^{2} +8\tilde{\lambda }\tilde{\eta }^{-1/2}\tilde{l}^{3}\Big),
\end{aligned}
\end{equation}
thus the range of the integral is determined.
In order to evaluate the integral, it is convenient to introduce the new variable $\sigma$ through
\begin{equation}\label{theta to sigma}
    \cos \theta =\cos \theta _{\mathrm{min}}\cos \sigma,
\end{equation}
thus the original integral over $\theta$ is transformed into the sum of integrals over $\sigma$, with the range being either from 0 to $\sigma_{\mathrm{s}}$ or from 0 to $\sigma_{\mathrm{o}}$.
\begin{equation}
\begin{aligned}
    &\ \ \ \ \int ^{\theta }\frac{|\mathrm{d\theta } |}{\sqrt{\Theta ( \theta )}} \\
    &=\left(\int _{\theta _{\mathrm{min}}}^{\theta _{\mathrm{s}}}+\int _{\theta _{\mathrm{min}}}^{\theta _{\mathrm{o}}}\right)
    \Big(\eta +( \lambda -a)^{2}
    -\frac{1}{\sin^{2} \theta }\left( \lambda -a\sin^{2} \theta +2l\cos \theta \right)^{2} \Big)^{-1/2} \mathrm{d\theta }\\
    &=\left(\int _{0}^{\sigma _{\mathrm{s}}} +\int _{0}^{\sigma _{\mathrm{o}}}\right)
    {r_{\mathrm{min}}^{(0)}}^{-1}
    \Big(1 - \tilde a^2 + 4\tilde l^2 
    + \tilde a^2 \cos^2 \theta_{\mathrm{min}}
    (\cos^2\sigma + 1)\\
    &\ \ \ \ +4 \tilde a \tilde l \cos\theta_{\mathrm{min}}
    \frac{\cos^2\sigma + \cos\sigma + 1}{\cos\sigma + 1}
    -(4\tilde a \tilde l-4\tilde\lambda \tilde l)\cos^{-1}\theta_{\mathrm{min}}\frac{1}{\cos\sigma + 1}
    \Big)^{-1/2}
    \mathrm{d} \sigma.
\end{aligned}
\end{equation}

By so far we have not inserted the value of $\theta_{\mathrm{min}}$ into the integrand above. After finishing this procedure and keeping the terms up to the third order of the small quantities,
we shall have
\begin{equation}
    \int ^{\theta }\frac{|\mathrm{d\theta } |}{\sqrt{\Theta ( \theta )}}
    =\left(\int _{0}^{\sigma _{\mathrm{s}}} +\int _{0}^{\sigma _{\mathrm{o}}}\right) H(\sigma) \mathrm{d}\sigma,
\end{equation}
where the full expression of $H(\sigma)$ is provided in equation (\ref{H sigma}) in the appendix.

Now we proceed to derive the polynomial equation of $\Delta\theta$, but before that we first express the integration as the polynomial of $\Delta\sigma$, since $\Delta\theta$ being small implies that
\begin{equation}
    \Delta\sigma = \sigma_{\mathrm{o}} + \sigma_{\mathrm{s}} - \pi
\end{equation}
is also a small quantity. Thus by denoting
\begin{equation}
    J(\sigma) = \int H(\sigma) \mathrm{d} \sigma,
\end{equation}
we have the Taylor expansion
\begin{equation}\label{Taylor expansion J}
\begin{aligned}
    \int ^{\theta }\frac{|\mathrm{d\theta } |}{\sqrt{\Theta ( \theta )}}
    &=J(\sigma_{\mathrm{s}})+J(\sigma_{\mathrm{o}})
    -2J(0)=J(\sigma_{\mathrm{s}})+J(\sigma_{\mathrm{o}})
    =J(\pi - \sigma_{\mathrm{o}} + \Delta\sigma )+J(\sigma_{\mathrm{o}})\\
    &= J(\pi - \sigma_{\mathrm{o}} )+J(\sigma_{\mathrm{o}})
    +J'(\pi - \sigma_{\mathrm{o}})\Delta\sigma
    +\frac{1}{2}J''(\pi - \sigma_{\mathrm{o}})\Delta\sigma^2\\
    &\ \ \ +\frac{1}{6}J'''(\pi - \sigma_{\mathrm{o}})\Delta\sigma^3
    +O(\Delta\sigma^4).
\end{aligned}
\end{equation}

This is the equation required for solving $\Delta\sigma$, since the value of its left hand side is determined by (\ref{spatial integration r1}), and the coefficients on the right hand side can also be calculated and are provided in equation (\ref{J coefficients}) the appendix.
Consequently, by denoting
\begin{equation}
\begin{aligned}
    \mathcal{L} &= \int ^{r}\frac{\mathrm{| d} r\mathrm{| }}{\sqrt{R( r)}} - (J(\pi - \sigma_{\mathrm{o}} )+J(\sigma_{\mathrm{o}})),\\
    \mathcal{M}_{1} &= J'(\pi - \sigma_{\mathrm{o}}),\\
    \mathcal{M}_{2} &= \frac{1}{2}J''(\pi - \sigma_{\mathrm{o}}),
\end{aligned}
\end{equation}
we have
\begin{equation}\label{L and M}
    \mathcal{L} = \mathcal{M}_{1}\Delta\sigma + \mathcal{M}_{2}\Delta\sigma^{2}
    +O(\Delta\sigma^4),
\end{equation}
where $\mathcal{L}$, $\mathcal{M}_{1}$ and $\mathcal{M}_{2}$ are all known values.
Therefore, by solving $\Delta\sigma$ from (\ref{L and M}), keeping the value which is a small quantity
\begin{equation}
    \Delta\sigma = \frac{-\mathcal{M}_{1}+\sqrt{\mathcal{M}_{1}^2+4\mathcal{M}_{2}\mathcal{L}}}{2\mathcal{M}_{2}}
\end{equation}
and expanding the result as the series of $\mathcal{L}$ up to the third order, we derive
\begin{equation}\label{delta sigma L}
    \Delta\sigma = \frac{1}{\mathcal{M}_{1}}\mathcal{L}
    -\frac{\mathcal{M}_{2}}{(\mathcal{M}_{1})^3}\mathcal{L}^2
    +\frac{2\mathcal{M}_{2}^2}{(\mathcal{M}_{1})^5}\mathcal{L}^3.
\end{equation}

The above result is transformed into the value of $\Delta\theta$ in the following approach.
With (\ref{theta to sigma}) we have
\begin{equation}
    \cos\theta_{\mathrm{s}} = \cos\theta_{\mathrm{min}}\cos\sigma_{\mathrm{s}},
\end{equation}
then through the Taylor expansion
\begin{equation}
\begin{aligned}
    \cos \theta _{\mathrm{s}} &=\cos( \pi -\theta _{\mathrm{o}} +\Delta \theta )\\
    &=-\cos \theta _{\mathrm{o}} -\sin \theta _{\mathrm{o}}  \Delta \theta +\frac{1}{2}\cos \theta _{\mathrm{o}}  \Delta \theta ^{2} +\frac{1}{6}\sin \theta _{\mathrm{o}}  \Delta \theta ^{3} +O\left( \Delta \theta ^{4}\right)
\end{aligned}
\end{equation}
and
\begin{equation}
\begin{aligned}
    \cos \sigma _{\mathrm{s}} &=\cos( \pi -\sigma _{\mathrm{o}} +\Delta \sigma )\\
    &=-\cos \sigma _{\mathrm{o}} -\sin \sigma _{\mathrm{o}}  \Delta \sigma +\frac{1}{2}\cos \sigma _{\mathrm{o}}  \Delta \sigma ^{2} +\frac{1}{6}\sin \sigma _{\mathrm{o}}  \Delta \sigma ^{3} +O\left( \Delta \sigma ^{4}\right),
\end{aligned}
\end{equation}
it holds up to the third order of $\Delta\theta$ and $\Delta\sigma$ that
\begin{equation}\label{equation delta theta delta sigma}
\begin{aligned}
    &\ \ \  -\sin \theta _{\mathrm{o}}  \Delta \theta +\frac{1}{2}\cos \theta _{\mathrm{o}}  \Delta \theta ^{2} +\frac{1}{6}\sin \theta _{\mathrm{o}}  \Delta \theta ^{3}\\
    &=\cos \theta _{\mathrm{min}}\left( -\sin \sigma _{\mathrm{o}}  \Delta \sigma +\frac{1}{2}\cos \sigma _{\mathrm{o}}  \Delta \sigma ^{2} +\frac{1}{6}\sin \sigma _{\mathrm{o}}  \Delta \sigma ^{3}\right).
\end{aligned}    
\end{equation}

From (\ref{equation delta theta delta sigma}) we derive
\begin{equation}\label{delta theta abc}
    \Delta \theta =\mathcal{A} \Delta \sigma +\mathcal{B} \Delta \sigma ^{2} +\mathcal{C} \Delta \sigma ^{3},
\end{equation}
where
\begin{equation}
\begin{aligned}
    \mathcal{A} &=\cos \theta _{\mathrm{min}}\frac{\sin \sigma _{\mathrm{o}}}{\sin \theta _{\mathrm{o}}}\\
    \mathcal{B} &=\frac{-\cos \theta _{\mathrm{min}}\cos \sigma _{\mathrm{o}}\sin^{2} \theta _{\mathrm{o}} +\cos^{2} \theta _{\mathrm{min}}\cos \theta _{\mathrm{o}}\sin^{2} \sigma _{\mathrm{o}}}{2\sin^{3} \theta _{\mathrm{o}}}\\
    \mathcal{C} &=-\frac{\sin \sigma _{\mathrm{o}}}{6\sin^{5} \theta _{\mathrm{o}}}( 3\cos^{2} \theta _{\mathrm{min}}\cos \theta _{\mathrm{o}}\cos \sigma _{\mathrm{o}}\sin^{2} \theta _{\mathrm{o}} +\sin^{4} \theta _{\mathrm{o}}\\
    &\ \ \ \ -3\cos^{3} \theta _{\mathrm{min}}\cos^{2} \theta _{\mathrm{o}}\sin^{2} \sigma _{\mathrm{o}} -\cos^{3} \theta _{\mathrm{min}}\sin^{2} \theta _{\mathrm{o}}\sin^{2} \sigma _{\mathrm{o}}).
\end{aligned}
\end{equation}
The final expression of $\Delta\theta$ is provided in the appendix.

The second integral of $\theta$ can be treated in the similar approach, in order to derive $\Delta\phi$ from (\ref{spatial integration}). By substitution of variable,
\begin{equation}\label{second integral of theta}
\begin{aligned}
    &\ \ \ \int ^{\theta }\frac{\left( \lambda \csc^{2} \theta +2l\csc^{2} \theta \cos \theta \right)\mathrm{| d\theta | }}{\sqrt{\Theta ( \theta )}}\\
    &=\left(\int _{\theta _{\mathrm{min}}}^{\theta _{\mathrm{s}}}+\int _{\theta _{\mathrm{min}}}^{\theta _{\mathrm{o}}}\right)
    \left( \lambda \csc^{2} \theta +2l\csc^{2} \theta \cos \theta \right)\\
    &\ \ \ \ \Big(\eta +( \lambda -a)^{2}
    -\frac{1}{\sin^{2} \theta }\left( \lambda -a\sin^{2} \theta +2l\cos \theta \right)^{2} \Big)^{-1/2} \mathrm{d\theta }\\
    &=\left(\int _{0}^{\sigma _{\mathrm{s}}} +\int _{0}^{\sigma _{\mathrm{o}}}\right)
    \left( \tilde\lambda +2\tilde l \cos \theta_{\mathrm{min}} \cos\sigma \right)
    \left(1- \cos^2 \theta_{\mathrm{min}} \cos^2\sigma \right)^{-1}\\
    &\ \ \ \ \Big(1 - \tilde a^2 + 4\tilde l^2 
    + \tilde a^2 \cos^2 \theta_{\mathrm{min}}
    (\cos^2\sigma + 1)
    +4 \tilde a \tilde l \cos\theta_{\mathrm{min}}
    \frac{\cos^2\sigma + \cos\sigma + 1}{\cos\sigma + 1}\\
    &\ \ \ \ -(4\tilde a \tilde l-4\tilde\lambda \tilde l)\cos^{-1}\theta_{\mathrm{min}}\frac{1}{\cos\sigma + 1}
    \Big)^{-1/2}
    \mathrm{d} \sigma.
\end{aligned}
\end{equation}
Then by denoting the primitive function of this integral as $P(\sigma)$,
\begin{equation}
\begin{aligned}
    &\ \ \ \int ^{\theta }\frac{\left( \lambda \csc^{2} \theta +2l\csc^{2} \theta \cos \theta \right)\mathrm{| d\theta | }}{\sqrt{\Theta ( \theta )}}\\
    &=P(\sigma_{\mathrm{s}})+P(\sigma_{\mathrm{o}})
    -2P(0)=P(\sigma_{\mathrm{s}})+P(\sigma_{\mathrm{o}})
    =P(\pi - \sigma_{\mathrm{o}} + \Delta\sigma )+P(\sigma_{\mathrm{o}})\\
    & = P(\pi - \sigma_{\mathrm{o}} )+P(\sigma_{\mathrm{o}})
    +P'(\pi - \sigma_{\mathrm{o}})\Delta\sigma
    +\frac{1}{2}P''(\pi - \sigma_{\mathrm{o}})\Delta\sigma^2\\
    &\ \ \ +\frac{1}{6}P'''(\pi - \sigma_{\mathrm{o}})\Delta\sigma^3
    +O(\Delta\sigma^4),
\end{aligned}    
\end{equation}
with the value of the coefficients in equation (\ref{P coefficients}) in the appendix. 

Since $\Delta\sigma$ is already known in (\ref{delta sigma L}), the right hand side of the second equation in (\ref{spatial integration}) is determined. Therefore, from (\ref{weak deflection}) we know that the value of $\Delta\phi$ can be calculated. The final result of $\Delta\phi$ is presented in the appendix.

Therefore, by inserting the expressions of $\Delta\theta$ and $\Delta\phi$ into $X$ and $x$ in (\ref{expansion X}) and (\ref{expansion x}), we finally arrive at the value of $\chi$, up to the third order of the small quantities, which is a simple expression depending on the physical parameters of KNTN space-time:
\begin{equation}
    \chi = \frac{a(5m^2 +Q^2 + 20l^2)\pi \cos\theta_{\mathrm{o}}}{4 r_{\mathrm{min}}^3}.
\end{equation}
When taking $Q=0$ and $l=0$, the above result goes back to the Faraday rotation in Kerr space-time \cite{Ishihara1988}. 

It is worth emphasising that the authors of \cite{Ishihara1988} did not pay much attention to the case where $\displaystyle\theta \equiv \frac{\pi}{2}$, and we should point out that in this case the geodesic equations should be rewritten from the beginning (see (\ref{S equatorial}-\ref{equation equatorial}) in the Appendix for details). Consequently, we cannot directly use the above result to derive $\chi = 0$. 
Instead, note that when $\displaystyle\theta \equiv \frac{\pi}{2}$, we have $\displaystyle k^{\theta} \equiv 0$, therefore the equation in (\ref{wave vector asymptotics}) leads to $\displaystyle\beta = 0$. Besides, from (\ref{beta and gamma}) it holds that $\displaystyle \gamma \equiv \lambda - a$ . Thus by plugging these equations into (\ref{original transformation matrix}), we derive $x=0$, or equivalently,
\begin{equation}\label{f transformation equatorial plane}
    \hat{f}^{\theta}_{\mathrm{o}}
    = \hat{f}^{\theta}_{\mathrm{s}},\
    \hat{f}^{\phi}_{\mathrm{o}}
    = -\hat{f}^{\phi}_{\mathrm{s}}.
\end{equation}
The first equation among them means that the component of the polarisation vector perpendicular to the equatorial plane remains the same for the source and the observer. Besides, when $\displaystyle\theta \equiv \frac{\pi}{2}$, the geodesic is lying in the equatorial plane, thus $\mathbf{n}$ is perpendicular to the equatorial plane by definition in (\ref{definition new frame}). Consequently, the $\mathbf{n}$-component of the polarisation vector is identical for the source and the observer. On the other hand, since (\ref{definition new frame}) implies that $\mathbf{h}_{\mathrm{s}}$ and $\mathbf{h}_{\mathrm{o}}$ both lie in the equatorial plane and share the same transformation relation with the second equation in (\ref{f transformation equatorial plane}), we are informed that the $\mathbf{h}$-component of polarisation vector is also identical for the source and the observer. As a result, there is no Faraday rotation. Readers can also check directly that the components of $\mathbf{n}$ satisfy $n^{\phi}=0$ and $n^{\theta} \neq 0$, such that when inserting into (\ref{definition of chi}), it follows that $X=0$ and $\chi = 0$. In conclusion, the gravitational Faraday rotation vanishes for geodesics lying in equatorial plane, even without the weak deflection assumption (\ref{weak deflection}).

\section{\label{discussions}Conclusions and discussions}

We calculated the Faraday rotation in Kerr-Newman-Taub-NUT space-time for remote source and observer. Compared with the vanishing result in \cite{Bini2003}, the result we obtained is non-zero and is dependent on the small quantities $\tilde{a}$, $\tilde{m}$, $\tilde{Q}$ and $\tilde{l}$. 
The calculation of rotation angle $\chi$ was based on the values $x$ and $X$, which are respectively derived from the Walker-Penrose constant and the choice of polarisation frame. We only considered the case with weak deflection of light ray, such that the deflection angles could be derived from the geodesic equations through Taylor expansion. 
However, in the scenario of strong deflection, it becomes less straightforward to use geodesic equations to establish the relationship between the constants of motion and the deflection angles. Therefore, we did not investigate this case, and numerical studies may be necessary for deeper understanding of this subject.

The method we adopted of choosing the polarisation basis for determining the Faraday rotation angle is not unique, and many other approaches exist in the literature. For instance, in \cite{Nouri-Zonoz1999} the null geodesic is projected onto the three-dimensional Riemannian space. In this projected space, the trajectory of light is no longer a geodesic, and neither is the projected polarisation vector parallelly propagated along the projected wave-vector. Instead, in this three-dimensional space the projected polarisation vector is rotating during the propagation, and the total rotation angle is recognised as the Faraday rotation. In other words, the parallel transported vectors in three-dimensional space are taken as the standard polarisation basis. This definition clearly differs from ours, and the result for the Kerr space-time turns out to be $\displaystyle\frac{a m^2\pi \cos\theta_{\mathrm{o}}}{4 r_{\mathrm{min}}^3}$, which is different from that in \cite{Ishihara1988}. We follow their procedure and calculated the result for KNTN space-time, which is $\displaystyle\frac{a(m^2 - Q^2 + l^2)\pi \cos\theta_{\mathrm{o}}}{4 r_{\mathrm{min}}^3}$ for weakly deflected light ray and remote source and observer. In \cite{Brodutch2011} the authors have chosen another polarisation basis which is neither based on the communication between the source and the observer (as in our work and \cite{Ishihara1988}), nor based on three-dimensional parallel transport (as in \cite{Nouri-Zonoz1999}), but is dependent on the free-fall acceleration direction. However, discussion of every existing definition of the polarisation basis and comparison between them is beyond the scope of this work, and we will leave this for future research.

\begin{acknowledgements}
The author is grateful to Xiaokai He, Xiaoning Wu and Naqing Xie for their valuable advice.
\end{acknowledgements}

\appendix*

\section{The geodesic equations and determination of the deflection angle}\label{spatial geodesic}

In this appendix, we mainly focus on the derivation of the geodesic equations and the usage of these equations to establish the relationship between constants of motion and the deflection angles. We start from the Hamilton-Jacobi equations for KNTN space-time and use separation of variables to derive ODEs of the coordinates. The case $\displaystyle \theta \equiv \frac{\pi}{2}$ is treated separately. Then we rewrite these equations in integral form and calculate the integrals over $r$ and $\theta$ respectively. Finally, we use Taylor expansion with respect to $\Delta\theta$ for the evaluation of these integrals, in order to solve the value of $\Delta\theta$ and $\Delta\phi$.

Following \cite{Chandrasekhar1983}, the Hamilton-Jacobi equation governing geodesic motion of KNTN metric
\begin{equation}
\begin{aligned}
    2 \frac{\partial S}{\partial \tau}
    &=g^{\mu \nu}\frac{\partial S}{\partial x^{\mu}}
    \frac{\partial S}{\partial x^{\nu}}\\
    &=-\frac{\Sigma}{\Delta}\left((1+\frac{a p}{\Sigma})
    \frac{\partial S}{\partial t}
    +\frac{a}{\Sigma}\frac{\partial S}{\partial \phi}\right)^2
    +\frac{\Sigma}{\sin^2\theta}
    \left(\frac{p}{\Sigma}\frac{\partial S}{\partial t}
    +\frac{1}{\Sigma}\frac{\partial S}{\partial \phi}\right)^2\\
    &\ \ \ +\frac{\Delta}{\Sigma}\left(\frac{\partial S}{\partial r}\right)^2
    +\frac{1}{\Sigma}\left( \frac{\partial S}{\partial \theta}\right)^2
\end{aligned}
\end{equation}
could be evaluated by imposing the assumption of the principal function for null geodesics
\begin{equation}
    S = -E t + L_{z} \phi + S_{r}(r) + S_{\theta}(\theta).
\end{equation}
 Here $E$ and $L_{z}$ are the conserved quantities corresponding to the Killing vectors $\displaystyle\frac{\partial}{\partial t}$ and $\displaystyle\frac{\partial}{\partial \phi}$ of the KNTN space-time:
\begin{equation}\label{conserve E Lz}
\begin{aligned}
    E &= \frac{1}{\Sigma}
    \left((\Delta-a^2\sin ^2\theta)\dot t
    +(\Delta p-a(\Sigma+a p)\sin^2\theta)\dot\phi\right)\\
    L_{z} &= \frac{1}{\Sigma}
    \left( (\Delta p-a(\Sigma+a p)\sin^2\theta)\dot t
    +((\Sigma+a p)^2\sin^2\theta - p^2 \Delta)\dot\phi\right),
\end{aligned}
\end{equation}
and the principal function $S$ does not explicitly include the affine parameter $\tau$ since the geodesics in consideration are null \cite{Carter1968}.

The resulting equation
\begin{equation}
    \frac{1}{\Delta}\left(-E(\Sigma+a p)+a L_{z}\right)^2
    -\frac{1}{\sin^2\theta}(-E p + L_{z})^2
    -\Delta\left(\frac{\mathrm{d}S_{r}}{\mathrm{d}r}\right)^2
    -\left(\frac{\mathrm{d}S_{\theta}}{\mathrm{d}\theta}\right)^2
    =0
\end{equation}
is separated with respect to the variables $r$ and $\theta$ into
\begin{equation}\label{separation of variables}
\begin{aligned}
    -\mathcal{Q}
    &=\Delta\left(\frac{\mathrm{d}S_{r}}{\mathrm{d}r}\right)^2
    -\frac{1}{\Delta}\left(-E (r^2+l^2+a^2)+a L_{z}\right)^2
    +(L_{z}-a E)^2,\\
    \mathcal{Q}
    &=\left(\frac{\mathrm{d}S_{\theta}}{\mathrm{d}\theta}\right)^2
    +\frac{1}{\sin^2\theta}
    (-p E+L_{z})^2-(L_{z}-a E)^2
\end{aligned}
\end{equation}
by introducing the Carter constant $\mathcal{Q}$.

Then since
\begin{equation}\label{generalised momentum}
\begin{aligned}
    \frac{\mathrm{d}S_{r}}{\mathrm{d}r}&\equiv p_{r}
    =\frac{\partial\mathscr{L}}{\partial \dot r}
    =\frac{\Sigma}{\Delta}\dot r,\\ 
    \frac{\mathrm{d}S_{\theta}}{\mathrm{d}\theta}&\equiv p_{\theta}
    =\frac{\partial\mathscr{L}}{\partial \dot \theta}
    =\Sigma\dot \theta,
\end{aligned}
\end{equation}
where
\begin{equation}
    \mathscr{L} = \frac{1}{2}g_{\mu\nu}\dot x^{\mu}\dot x^{\nu}
\end{equation}
is the Lagrangian of KNTN space-time satisfying
\begin{equation}
    S = \int \mathscr{L} d\tau,
\end{equation}
the $r$ and $\theta$ components of geodesic equations turn out to be
\begin{equation}\label{r theta geodesic}
\begin{aligned}
    \Sigma \dot r &= \pm \sqrt{R(r)}\\
    \Sigma \dot \theta &= \pm \sqrt{\Theta(\theta)},
\end{aligned}
\end{equation}
where
\begin{equation}
\begin{aligned}
    R( r) &=-\Delta \left( \mathcal{Q} +( L_{z} -a E)^{2}\right)
    +\left( a L_{z} -E \left( r^{2} +l^{2} +a^{2}\right)\right)^{2},\\
    \Theta ( \theta ) &=\mathcal{Q} +( L_{z} -a E)^{2}
    -\frac{1}{\sin^{2} \theta }\left(- E (a\sin^{2} \theta -2l\cos \theta)+ L_{z} \right)^{2}.
\end{aligned}
\end{equation}
In the following part we will take $E=1$, $\displaystyle\lambda = \frac{L_{z}}{E}=L_{z}$ and $\displaystyle\eta = \frac{\mathcal{Q}}{E^2}=\mathcal{Q}$ for simplicity.

However, the above procedure of variable separation does not apply to the geodesics with $\displaystyle \theta \equiv \frac{\pi}{2}$, i.e., those lying in the orbital plane. Instead, by inserting the assumption
\begin{equation}\label{S equatorial}
    S = -E t + L_{z} \phi + S_{r}(r)
\end{equation}
into
\begin{equation}
\begin{aligned}
    2 \frac{\partial S}{\partial \tau}
    &=-\frac{\Sigma}{\Delta}\left((1+\frac{a p}{\Sigma})
    \frac{\partial S}{\partial t}
    +\frac{a}{\Sigma}\frac{\partial S}{\partial \phi}\right)^2
    +\Sigma
    \left(\frac{p}{\Sigma}\frac{\partial S}{\partial t}
    +\frac{1}{\Sigma}\frac{\partial S}{\partial \phi}\right)^2
    +\frac{\Delta}{\Sigma}\left(\frac{\partial S}{\partial r}\right)^2,
\end{aligned}
\end{equation}
we derive
\begin{equation}\label{equation equatorial}
    \frac{1}{\Delta}\left(-E(\Sigma+a p)+a L_{z}\right)^2
    -(-E p + L_{z})^2
    -\Delta\left(\frac{\mathrm{d}S_{r}}{\mathrm{d}r}\right)^2
    =0,
\end{equation}
which is the $r$ component of geodesic equations.
Note that (\ref{separation of variables}) and (\ref{generalised momentum}) still hold in this case as long as we set $\mathcal{Q}=0$ and $S_{\theta}$ being any constant. Therefore the general geodesic equations (\ref{r theta geodesic}) are not violated, with $\eta = 0$ and $\Theta \equiv 0$.

The remaining two equations are transformed from (\ref{conserve E Lz}):
\begin{equation}
\begin{aligned}
    \dot t &=\frac{1}{\Sigma \Delta }\left( r^{2} +a^{2} +l^{2}\right)\left( r^{2} +a^{2} +l^{2} -a\lambda \right) \\
    &\ \ \ +\frac{1}{\Sigma \sin^{2} \theta }\left( a\sin^{2} \theta -2l\cos \theta \right)\left( \lambda -a\sin^{2} \theta +2l\cos \theta \right),\\
    \dot\phi &=\frac{a}{\Sigma \Delta }\left( r^{2} +a^{2} +l^{2} -a\lambda \right)
    +\frac{1}{\Sigma \sin^{2} \theta }\left( \lambda -a\sin^{2} \theta +2l\cos \theta \right) .
\end{aligned}
\end{equation}

In the following we will provide some details in the treatment of the spatial geodesic equations in the integral form.
In order to evaluate the integral
\begin{equation}
    \int _{r_{\mathrm{min}}}^{\infty }\frac{\mathrm{d} r}{\sqrt{R( r)}},
\end{equation}
we utilise the techniques in \cite{Bray1986}, 
introducing the new dimensionless variable
\begin{equation}\label{substitution of new variable}
    \mathbf{x} = \frac{r_{\mathrm{min}}}{r}
\end{equation}
such that
\begin{equation}\label{substitution of new variable integration}
\begin{aligned}
    &\ \ \ \ 2\int _{r_{\mathrm{min}}}^{\infty }\frac{\mathrm{d} r}{\sqrt{R( r)}}\\
    &=2\int_{0}^{1}\Big(
    r_{\mathrm{min}}^2
    +(a^2+2l^2-\lambda^2-\eta)\mathbf{x}^2
    +2m\left(\eta+(\lambda-a)^2\right)\mathbf{x}^3/r_{\mathrm{min}}\\
    &\ \ \ +\left((a^2+l^2-a \lambda)^2+(l^2-a^2-Q^2)(\eta+(\lambda-a)^2)\right)\mathbf{x}^4/r_{\mathrm{min}}^2
    \Big)^{-1/2}\mathrm{d}\mathbf{x}\\
    &=2\int_{0}^{1}\Big(
    (a^2+2l^2-\lambda^2-\eta)(\mathbf{x}^2-1)
    +2m\left(\eta+(\lambda-a)^2\right)(\mathbf{x}^3-1)/r_{\mathrm{min}}\\
    &\ \ \ +\left((a^2+l^2-a \lambda)^2+(l^2-a^2-Q^2)(\eta+(\lambda-a)^2)\right)(\mathbf{x}^4-1)/r_{\mathrm{min}}^2
    \Big)^{-1/2}\mathrm{d}\mathbf{x}\\
    &=2\int_{0}^{1}(r_{\mathrm{min}}^{(0)})^{-1}
    (1-\mathbf{x}^2)^{-1/2}
    \Bigg(
    1-\tilde a^2-2\tilde l^2
    -2\tilde m\big(\tilde\eta + (\tilde \lambda-\tilde a)^2\big)
    \frac{r_{\mathrm{min}}^{(0)}}{r_{\mathrm{min}}}
    \frac{\mathbf{x}^2+\mathbf{x}+1}{\mathbf{x}+1}\\
    &\ \ \ -\Big((\tilde a^2+\tilde l^2-\tilde a \tilde\lambda)^2+(\tilde l^2-\tilde a^2-\tilde Q^2)
    \big(\tilde \eta+(\tilde\lambda-\tilde a)^2\big)\Big)
    \Big(\frac{r_{\mathrm{min}}^{(0)}}{r_{\mathrm{min}}}\Big)^2
    (\mathbf{x}^2+1)
    \Bigg)^{-1/2}\mathrm{d}\mathbf{x}.
\end{aligned}
\end{equation}
After Taylor expansion of the integrand up to the third order of $\tilde a$, $\tilde m$, $\tilde Q$ and $\tilde l$, the resulting function can be directly integrated over $\mathbf{x}$ as in \cite{Bray1986}, such that we have
\begin{equation}
\begin{aligned}
    \int ^{r}\frac{\mathrm{| d} r\mathrm{| }}{\sqrt{R( r)}} 
    &= r{_{\mathrm{min}}^{( 0)}}^{-1}
    \Big(
     \pi+4\tilde{m}+32\tilde{m}\tilde{l}^{2} -16\tilde{m}\tilde{Q}^{2} -8\tilde{m}\tilde{a}\tilde{\lambda } 
    +4\tilde{m}\tilde{a}^{2}( 4\tilde{\lambda }^{2} -1)\\
    &\ \ \ +\frac{7}{4}\pi \tilde{l}^{2} 
    -\frac{3}{4} \pi \tilde{Q}^{2}
    -3\pi\tilde{a}\tilde{l}^{2}\tilde{\lambda }
    +\frac{3}{2}\pi\tilde{a}\tilde{Q}^{2}\tilde{\lambda }
    -\frac{1}{4}\pi\tilde{a}^{2}
    +\frac{3}{4}\pi\tilde{\lambda }^{2}\tilde{a}^{2}\\
    &\ \ \ +\frac{128}{3}\tilde{m}^{3} +\frac{15}{4} \pi \tilde{m}^{2} -15\pi \tilde{m}^{2}\tilde{a}\tilde{\lambda }\Big),
\end{aligned}    
\end{equation}
and the other integral involving $r$ can be treated in the same manner.

For the integration involving $\theta$, 
we have derived
\begin{equation}
    \int ^{\theta }\frac{|\mathrm{d\theta } |}{\sqrt{\Theta ( \theta )}}
    =\left(\int _{0}^{\sigma _{\mathrm{s}}} +\int _{0}^{\sigma _{\mathrm{o}}}\right) H(\sigma) \mathrm{d}\sigma,
\end{equation}
where
\begin{equation}\label{H sigma}
\begin{aligned}
    H(\sigma) &= {r_{\mathrm{min}}^{(0)}}^{-1}
    \Bigg( 
    1-\frac{2\tilde{\lambda }\tilde{\eta }^{-1/2}}{\cos \sigma +1}\tilde{l} +2\tilde{\eta }^{-1/2}\big(\frac{\tilde{\lambda }^{2}}{\cos \sigma +1} -\tilde{\eta }\cos \sigma \big)\tilde{a}\tilde{l}\\
    &\ \ \ +\frac{1}{2}(\tilde{\lambda }^{2} -\tilde{\eta }\cos^{2} \sigma )\tilde{a}^{2} 
    -\frac{2\big( 1-\tilde{\lambda }^{2}\tilde{\eta }^{-1} +2( 1+\tilde{\lambda }^{2}\tilde{\eta }^{-1})\cos \sigma +\cos^{2} \sigma \big)}{(\cos \sigma +1)^{2}}\tilde{l}^{2}\\
    &\ \ \ +\frac{\tilde{\lambda }\tilde{\eta }^{-1/2}\big( 2(\tilde{\eta } -\tilde{\lambda }^{2}) +\tilde{\eta }\cos \sigma (5\cos \sigma +\cos 2\sigma +3)\big)}{\cos \sigma +1}\tilde{a}^{2}\tilde{l}\\
    &\ \ \ +\frac{\tilde{\lambda }\big( 18-4\tilde{\lambda }^{2}\tilde{\eta }^{-1} +( 27+8\tilde{\lambda }^{2}\tilde{\eta }^{-1})\cos \sigma +10\cos 2\sigma +\cos 3\sigma \big)}{(\cos \sigma +1)^{2}}\tilde{a}\tilde{l}^{2}\\
    &\ \ \ +\frac{2\tilde{\lambda }\tilde{\eta }^{-1/2}\big( 6-\tilde{\lambda }^{2}\tilde{\eta }^{-1} +8( 1+\tilde{\lambda }^{2}\tilde{\eta }^{-1})\cos \sigma +( 2-\tilde{\lambda }^{2}\tilde{\eta }^{-1})\cos 2\sigma \big)}{(\cos \sigma +1)^{3}}\tilde{l}^{3}
    \Bigg).
\end{aligned}
\end{equation}

After denoting
\begin{equation}
    J(\sigma) = \int H(\sigma) \mathrm{d} \sigma,
\end{equation}
it leads to
\begin{equation}\label{J coefficients}
\begin{aligned}
    J(\pi - \sigma_{\mathrm{o}} )+J(\sigma_{\mathrm{o}})
    &= r{_{\mathrm{min}}^{( 0)}}^{-1}
    \Big(
    \pi -4\tilde{\lambda }\tilde{\eta }^{-1/2}\csc \sigma _{\mathrm{o}}\tilde{l} 
    -\frac{1}{4} \pi (\tilde{\eta } -2\tilde{\lambda }^{2})\tilde{a}^{2}\\
    &\ \ \ +4\tilde{\eta }^{-1/2}(\tilde{\lambda }^{2}\csc \sigma _{\mathrm{o}} -4\tilde{\eta }\sin \sigma _{\mathrm{o}})\tilde{a}\tilde{l}\\
    &\ \ \ -2\tilde{\eta }^{-1}( \pi \tilde{\eta } -4\tilde{\lambda }^{2}\cot^{2} \sigma _{\mathrm{o}}\csc \sigma _{\mathrm{o}})\tilde{l}^{2}\\
    &\ \ \ +\tilde{\lambda }\tilde{\eta }^{-1/2}\csc \sigma _{\mathrm{o}}( 9\tilde{\eta } -4\tilde{\lambda }^{2} -3\tilde{\eta }\cos 2\sigma _{\mathrm{o}})\tilde{a}^{2}\tilde{l}\\
    &\ \ \ +4\tilde{\lambda }\tilde{\eta }^{-1}\big( 3\pi \tilde{\eta } +\cot^{2} \sigma _{\mathrm{o}}\csc \sigma _{\mathrm{o}}( -\tilde{\eta } -4\tilde{\lambda }^{2} +\tilde{\eta }\cos 2\sigma _{\mathrm{o}})\big)\tilde{a}\tilde{l}^{2}\\
    &\ \ \ +\frac{8}{3}\tilde{\lambda }\tilde{\eta }^{-3/2}\csc \sigma _{\mathrm{o}}\big( 6\tilde{\eta } -3\tilde{\lambda }^{2} +\tilde{\lambda }^{2}\csc^{2} \sigma _{\mathrm{o}}( 17-12\csc^{2} \sigma _{\mathrm{o}})\big)\tilde{l}^{3}
    \Big),\\
    J'(\pi - \sigma_{\mathrm{o}})
    &= r{_{\mathrm{min}}^{( 0)}}^{-1}
    \Big(
    1-\tilde{\lambda }\tilde{\eta }^{-1/2}\csc^{2}\frac{\sigma _{\mathrm{o}}}{2}\tilde{l} 
    -\frac{1}{4}(\tilde{\eta } -2\tilde{\lambda }^{2} +\tilde{\eta }\cos 2\sigma _{\mathrm{o}})\tilde{a}^{2}\\
    &\ \ \ +\tilde{\eta }^{-1/2}( 2\tilde{\eta }\cos \sigma _{\mathrm{o}} +\tilde{\lambda }^{2}\csc^{2}\frac{\sigma _{\mathrm{o}}}{2})\tilde{a}\tilde{l}\\
    &\ \ \ +\frac{1}{2}\tilde{\eta }^{-1}\big(\tilde{\lambda }^{2} (2\cos \sigma _{\mathrm{o}} +1)\csc^{4}\frac{\sigma _{\mathrm{o}}}{2} -4\tilde{\eta }\big)\tilde{l}^{2}
    \Big),\\
    \frac{1}{2}J''(\pi - \sigma_{\mathrm{o}})
    &= r{_{\mathrm{min}}^{( 0)}}^{-1}
    \Big(-\frac{1}{4}\tilde{\lambda }\tilde{\eta }^{-1/2}\sin \sigma _{\mathrm{o}}\csc^{4}\frac{\sigma _{\mathrm{o}}}{2}\tilde{l}\Big),\\
    \frac{1}{6}J'''(\pi - \sigma_{\mathrm{o}})
    &= 0,
\end{aligned}
\end{equation}
where we have kept the terms up to the third order for 
$\displaystyle J(\pi - \sigma_{\mathrm{o}} )+J(\sigma_{\mathrm{o}})$, second order for
$\displaystyle J'(\pi - \sigma_{\mathrm{o}})$, first order for
$\displaystyle \frac{1}{2}J''(\pi - \sigma_{\mathrm{o}})$ and zeroth order for $\displaystyle \frac{1}{6}J'''(\pi - \sigma_{\mathrm{o}})$, such that Taylor expansion up to the third order are considered overall in (\ref{Taylor expansion J}).

After the calculation of $\Delta\sigma$ through (\ref{delta sigma L}), the value of $\Delta\theta$ can be derived through (\ref{delta theta abc}), up to the third order of $\tilde{m}$, $\tilde{a}$, $\tilde{Q}$ and $\tilde{l}$. The result can be expressed in the form
\begin{equation}
\begin{aligned}
    \Delta \theta &=C_{m}\tilde{m} +C_{a}\tilde{a} +C_{Q}\tilde{Q} +C_{l}\tilde{l}
    +C_{mm}\tilde{m}^{2} +C_{ma}\tilde{m}\tilde{a} +C_{mQ}\tilde{m}\tilde{Q} +C_{ml}\tilde{m}\tilde{l}\\
    &\ \ \ +C_{aa}\tilde{a}^{2} +C_{aQ}\tilde{a}\tilde{Q} +C_{al}\tilde{a}\tilde{l} +C_{QQ}\tilde{Q}^{2} +C_{Ql}\tilde{Q}\tilde{l} +C_{ll}\tilde{l}^{2}\\
    &\ \ \ +C_{mmm}\tilde{m}^{3} +C_{mma}\tilde{m}^{2}\tilde{a} +C_{mmQ}\tilde{m}^{2}\tilde{Q} +C_{mml}\tilde{m}^{2}\tilde{l}\\
    &\ \ \ +C_{maa}\tilde{m}\tilde{a}^{2} +C_{maQ}\tilde{m}\tilde{a}\tilde{Q} +C_{mal}\tilde{m}\tilde{a}\tilde{l} +C_{mQQ}\tilde{m}\tilde{Q}^{2} +C_{mQl}\tilde{m}\tilde{Q}\tilde{l} +C_{mll}\tilde{m}\tilde{l}^{2}\\
    &\ \ \ +C_{aaa}\tilde{a}^{3} +C_{aaQ}\tilde{a}^{2}\tilde{Q} +C_{aal}\tilde{a}^{2}\tilde{l} +C_{aQQ}\tilde{a}\tilde{Q}^{2} +C_{aQl}\tilde{a}\tilde{Q}\tilde{l} +C_{all}\tilde{a}\tilde{l}^{2}\\
    &\ \ \ +C_{QQQ}\tilde{Q}^{3} +C_{QQl}\tilde{Q}^{2}\tilde{l} +C_{Qll}\tilde{Q}\tilde{l}^{2} +C_{lll}\tilde{l}^{2},
\end{aligned}
\end{equation}
with the following coefficients for the first order:
\begin{equation}
C_{m} = 4\csc \theta _{\mathrm{o}} \mu,\
C_{l} = 4\tilde{\lambda }\csc \theta _{\mathrm{o}},\
C_{a} = C_{Q}=0,
\end{equation}
the second order:
\begin{equation}
\begin{aligned}
&C_{mm} = \frac{15}{4} \pi \csc \theta _{\mathrm{o}} \mu -8\tilde{\lambda }^{2}\cot \theta _{\mathrm{o}}\csc^{2} \theta _{\mathrm{o}},\
C_{ma} = -8\tilde{\lambda }\csc \theta _{\mathrm{o}} \mu,\\
&C_{ml} = 8\tilde{\lambda }\cot \theta _{\mathrm{o}}(1 -2\csc^{2} \theta _{\mathrm{o}}\tilde{\lambda }^{2} ) \mu ^{-1},\
C_{al} = 4\sin \theta _{\mathrm{o}} -8\tilde{\lambda }^{2}\csc \theta _{\mathrm{o}},\\
&C_{QQ} = -\frac{3}{4} \pi \csc \theta _{\mathrm{o}} \mu,\
C_{ll} = 8\tilde{\lambda }^{2}\cot \theta _{\mathrm{o}}\csc^{2} \theta _{\mathrm{o}},\\
&C_{mQ} = C_{aa} = C_{aQ} = C_{Ql} = 0,
\end{aligned}
\end{equation}
and the third order:
\begin{equation}
\begin{aligned}
&C_{mmm} = -\frac{32}{3}\tilde{\lambda }^{2} (\cos 2\theta _{\mathrm{o}} +2)\csc^{5} \theta _{\mathrm{o}} \mu -15\pi \tilde{\lambda }^{2}\cot \theta _{\mathrm{o}}\csc^{2} \theta _{\mathrm{o}} +\frac{128}{3}\csc \theta _{\mathrm{o}} \mu ,\\
&C_{mma} = \tilde{\lambda }\csc \theta _{\mathrm{o}}( 32\tilde{\lambda }^{2}\cot \theta _{\mathrm{o}}\csc \theta _{\mathrm{o}} -15\pi \mu ),\\
&C_{mml} = 16 \tilde{\lambda }\cos \theta _{\mathrm{o}}\cot \theta _{\mathrm{o}} 
+ 96\tilde{\lambda }^{3}\csc \theta _{\mathrm{o}}\cot^{4} \theta _{\mathrm{o}} \mu ^{-2}\\
&\ \ \ \ \ \ \ \ \  -\frac{15}{2}\pi \tilde{\lambda }\csc \theta _{\mathrm{o}}(\cos\theta_{\mathrm{o}}-2\tilde\lambda^2\cot\theta_{\mathrm{o}}\csc\theta_{\mathrm{o}}) \mu^{-1} \\
&\ \ \ \ \ \ \ \ \ +128\tilde{\eta }\tilde{\lambda }\csc \theta _{\mathrm{o}}\big( -2\tilde{\lambda }^{2} (\cos 2\theta _{\mathrm{o}} +2)\csc^{4} \theta _{\mathrm{o}} +1\big)\mu ^{-2},\\
&C_{maa} = 2\tilde{\eta }\csc \theta _{\mathrm{o}}\tilde{\lambda }^{2} \mu ^{-1} +( 11\tilde{\lambda }^{2} -3\tilde{\eta } +\cos 2\theta _{\mathrm{o}})\csc \theta _{\mathrm{o}} \mu ,\\
&C_{mal} =\big(\sin 4\theta _{\mathrm{o}} -2( 7\tilde{\lambda }^{2} +\tilde{\eta })\sin 2\theta _{\mathrm{o}}\big)\mu ^{-3} +32\cot \theta _{\mathrm{o}}( 3\tilde{\lambda }^{4} -2\tilde{\lambda }^{6}\csc^{2} \theta _{\mathrm{o}}) \mu ^{-3},\\
&C_{mQQ} = ( 3\pi \tilde{\lambda }^{2}\cot \theta _{\mathrm{o}}\csc \theta _{\mathrm{o}} -16\mu )\csc \theta _{\mathrm{o}} ,\\
&C_{mll} = -32\tilde{\lambda }^{4} (\cos 2\theta _{\mathrm{o}} +2)\csc^{5} \theta _{\mathrm{o}} \mu ^{-1} +40\csc \theta _{\mathrm{o}} \mu -64\csc^{3} \theta _{\mathrm{o}}\tilde{\lambda }^{4}\mu ^{-3}\\
&\ \ \ \ \ \ -8( 11\tilde{\lambda }^{2} +8\tilde{\eta })\csc \theta _{\mathrm{o}}\tilde{\lambda }^{2}\mu ^{-3}
+8( 3\tilde{\lambda }^{2} +\tilde{\eta })\sin \theta _{\mathrm{o}}\mu ^{-3},\\
&C_{aal} = 4\tilde{\lambda }( 2\tilde{\lambda }^{2} -2\tilde{\eta } +\cos 2\theta _{\mathrm{o}})\csc \theta _{\mathrm{o}},\\
&C_{aQQ} = \frac{3}{2} \pi \tilde{\lambda }\csc \theta _{\mathrm{o}} \mu,\\
&C_{all} = -\tilde{\lambda }\csc \theta _{\mathrm{o}}( 32\cot \theta _{\mathrm{o}}\csc \theta _{\mathrm{o}}\tilde{\lambda }^{2} +15\pi \mu ) ,\\
&C_{QQl} = -\frac{3}{4}\pi \tilde{\lambda }( 4\cot \theta _{\mathrm{o}}\csc^{2} \theta _{\mathrm{o}}\tilde{\lambda }^{4} -6\cot \theta _{\mathrm{o}}\tilde{\lambda }^{2} +\sin 2\theta _{\mathrm{o}}) \mu ^{-3},\\
&C_{lll} = -\frac{16}{3}\tilde{\lambda }\big( 3-2\tilde{\lambda }^{2} (\cos 2\theta _{\mathrm{o}} +2)\csc^{4} \theta _{\mathrm{o}}\big)\csc \theta _{\mathrm{o}} \\
&\ \ \ \ \ \ \ +\frac{15}{2} \pi \tilde{\lambda }(\sqrt{\tilde{\eta }} +\cos \theta _{\mathrm{o}}) \mu ^{-1}\csc \theta _{\mathrm{o}} +\frac{15}{4} \pi \csc \theta _{\mathrm{o}} \mu ,\\
&C_{mmQ} = C_{maQ} = C_{mQl} = C_{aaa} = C_{aaQ} = C_{aQl} = C_{QQQ} = C_{Qll} = 0.
\end{aligned}
\end{equation}
Note that in the above expressions we have $\displaystyle\tilde\eta = 1-\tilde\lambda^2$ by definition (\ref{tilde notations}). Thus by taking the sum of all these terms, the expression of $\Delta\theta$ is derived. 

In the treatment of the second integral over theta (\ref{second integral of theta}), we introduced the primitive function $P(\sigma)$. The following values are calculated where we kept the terms up to the third order for 
$\displaystyle P(\pi - \sigma_{\mathrm{o}} )+P(\sigma_{\mathrm{o}})$, second order for
$\displaystyle P'(\pi - \sigma_{\mathrm{o}})$, first order for
$\displaystyle \frac{1}{2}P''(\pi - \sigma_{\mathrm{o}})$ and zeroth order for $\displaystyle \frac{1}{6}P'''(\pi - \sigma_{\mathrm{o}})$ with respect to the small quantities:
\begin{equation}\label{P coefficients}
\begin{aligned}
    P(\pi - \sigma_{\mathrm{o}} )+P(\sigma_{\mathrm{o}})
    &=-4\mu ^{-1}\tilde{l} +\frac{1}{2} \pi \tilde{\lambda }\tilde{a}^{2} +4\tilde{\lambda } \mu ^{-1}\tilde{a}\tilde{l}\\
    &\ \ \ +2\left(\cos^{2} \theta _{\mathrm{o}}\tilde{\lambda }^{2} \mu ^{-3} +\left( 2\tilde{\eta } -2\tilde{\lambda }^{2} -\cos^{2} \theta _{\mathrm{o}}\right) \mu ^{-1}\right)\tilde{a}^{2}\tilde{l}\\
    &\ \ \ +4\pi \tilde{a}\tilde{l}^{2} +\frac{8}{3}\left( 2\tilde{\eta } -\cos^{2} \theta _{\mathrm{o}}\left( 2+3\sin^{2} \theta _{\mathrm{o}}\right)\right) \mu ^{-5}\tilde{l}^{3},\\
    P'(\pi - \sigma_{\mathrm{o}})
    &=\tilde{\lambda }\csc^{2} \theta _{\mathrm{o}} -2\left(\cos \theta _{\mathrm{o}} +\tilde{\lambda }^{2}\tilde{\eta }^{1/2}\csc^{2} \theta _{\mathrm{o}}\right) \mu ^{-2}\tilde{l}\\
    &\ \ \ +\left( 4\tilde{\eta }^{1/2}\cos \theta _{\mathrm{o}} +3\cos 2\theta _{\mathrm{o}} +\left(\tilde{\eta } +7\tilde{\lambda }^{2}\right) -4\tilde{\lambda }^{4}\csc^{2} \theta _{\mathrm{o}}\right)\tilde{\lambda } \mu ^{-4}\tilde{l}^{2}\\
    &\ \ \ +\frac{1}{2}\left( -\tilde{\eta }\csc^{2} \theta _{\mathrm{o}} +1\right)\tilde{\lambda }\tilde{a}^{2} +2\left(\tilde{\lambda }^{2}\tilde{\eta }^{1/2}\csc^{2} \theta _{\mathrm{o}} +\cos \theta _{\mathrm{o}}\right) \mu ^{-2}\tilde{\lambda }\tilde{a}\tilde{l},\\
    \frac{1}{2}P''(\pi - \sigma_{\mathrm{o}})
    &=\tilde{\lambda }\cot \theta _{\mathrm{o}}\csc^{3} \theta _{\mathrm{o}} \mu 
    -\Big((\cos \theta _{\mathrm{o}} +\tilde{\eta }^{1/2})^{2} \mu ^{-3} \\
    &\ \ \ \ \ \ +\cot \theta _{\mathrm{o}}\csc^{3} \theta _{\mathrm{o}}(\cos 2\theta _{\mathrm{o}} -\tilde{\eta } +3\tilde{\lambda }^{2})\tilde{\eta }^{1/2} \mu ^{-1}\Big)\tilde{l},\\
    \frac{1}{6}P'''(\pi - \sigma_{\mathrm{o}})
    &=\frac{1}{3}\tilde{\lambda }\csc^{2} \theta _{\mathrm{o}}\left( 3\left(\tilde{\eta } +2\tilde{\lambda }^{2}\right)\csc^{2} \theta _{\mathrm{o}} -2-4\tilde{\lambda }^{2}\csc^{4} \theta _{\mathrm{o}}\right).
\end{aligned}
\end{equation}

With these coefficients and other quantities already obtained, the value of $\Delta\phi$ can be determined by plugging in the expression of $\Delta \sigma$ into (\ref{spatial integration}), and keeping the terms up to the third order of $\tilde a$, $\tilde m$, $\tilde Q$ and $\tilde l$. Finally we arrive at
\begin{equation}
\begin{aligned}
    \Delta \phi &=F_{m}\tilde{m} +F_{a}\tilde{a} +F_{Q}\tilde{Q} +F_{l}\tilde{l}
    +F_{mm}\tilde{m}^{2} +F_{ma}\tilde{m}\tilde{a} +F_{mQ}\tilde{m}\tilde{Q} +F_{ml}\tilde{m}\tilde{l}\\
    &\ \ \ +F_{aa}\tilde{a}^{2} +F_{aQ}\tilde{a}\tilde{Q} +F_{al}\tilde{a}\tilde{l} +F_{QQ}\tilde{Q}^{2} +F_{Ql}\tilde{Q}\tilde{l} +F_{ll}\tilde{l}^{2}\\
    &\ \ \ +F_{mmm}\tilde{m}^{3} +F_{mma}\tilde{m}^{2}\tilde{a} +F_{mmQ}\tilde{m}^{2}\tilde{Q} +F_{mml}\tilde{m}^{2}\tilde{l}\\
    &\ \ \ +F_{maa}\tilde{m}\tilde{a}^{2} +F_{maQ}\tilde{m}\tilde{a}\tilde{Q} +F_{mal}\tilde{m}\tilde{a}\tilde{l} +F_{mQQ}\tilde{m}\tilde{Q}^{2} +F_{mQl}\tilde{m}\tilde{Q}\tilde{l} +F_{mll}\tilde{m}\tilde{l}^{2}\\
    &\ \ \ +F_{aaa}\tilde{a}^{3} +F_{aaQ}\tilde{a}^{2}\tilde{Q} +F_{aal}\tilde{a}^{2}\tilde{l} +F_{aQQ}\tilde{a}\tilde{Q}^{2} +F_{aQl}\tilde{a}\tilde{Q}\tilde{l} +F_{all}\tilde{a}\tilde{l}^{2}\\
    &\ \ \ +F_{QQQ}\tilde{Q}^{3} +F_{QQl}\tilde{Q}^{2}\tilde{l} +F_{Qll}\tilde{Q}\tilde{l}^{2} +F_{lll}\tilde{l}^{2},
\end{aligned}
\end{equation}
with the first-order coefficients:
\begin{equation}
    F_{m} = 4\tilde{\lambda }\csc^{2} \theta _{\mathrm{o}} ,\
    F_{l} = -4(1 -\csc^{2} \theta _{\mathrm{o}}\tilde{\lambda }^{2} ) \mu ^{-1},\ 
    F_{a} = F_{Q} = 0,
\end{equation}
the second-order coefficients:
\begin{equation}
\begin{aligned}
&F_{mm} = 16\tilde{\lambda }\cot \theta _{\mathrm{o}}\csc^{3} \theta _{\mathrm{o}} \mu +\frac{15}{4} \pi \tilde{\lambda }\csc^{2} \theta _{\mathrm{o}},\
F_{ma} = 4-8\tilde{\lambda }^{2}\csc^{2} \theta _{\mathrm{o}},\\
&F_{ml} = 8\cot \theta _{\mathrm{o}}\csc \theta _{\mathrm{o}}( 4\tilde{\lambda }^{2}\csc^{2} \theta _{\mathrm{o}} -1 ) ,\ 
F_{al} = 8\tilde{\lambda }(1 -\csc^{2} \theta _{\mathrm{o}}\tilde{\lambda }^{2} ) \mu ^{-1},\\
&F_{QQ} = -\frac{3}{4}\pi \tilde{\lambda }\csc^{2} \theta _{\mathrm{o}},\
F_{ll} = -8\tilde{\lambda }\big(\cot \theta _{\mathrm{o}}\csc \theta _{\mathrm{o}}( 2\tilde{\lambda }^{2}\csc^{2} \theta _{\mathrm{o}} -3)\tilde{\lambda }^{2} +\cos \theta _{\mathrm{o}}\big) \mu ^{-3},\\
&F_{mQ} = F_{aa} = F_{aQ} = F_{Ql} = 0,
\end{aligned}
\end{equation}
and the third-order coefficients:
\begin{equation}
\begin{aligned}
&F_{mmm} =  30\pi \tilde{\lambda }\csc^{3} \theta _{\mathrm{o}}\cot \theta _{\mathrm{o}} \mu 
-\frac{256}{3}\tilde{\lambda }^{3}\csc^{6} \theta _{\mathrm{o}} +64\tilde{\lambda }( 2\tilde{\lambda }^{2} +\tilde{\eta })\csc ^{4}\theta _{\mathrm{o}}, \\
&F_{mma} = 5\pi  -64\tilde{\lambda }^{2}\cot \theta _{\mathrm{o}}\csc^{3} \theta _{\mathrm{o}} \mu -15\pi \tilde{\lambda }^{2}\csc^{2} \theta _{\mathrm{o}}, \\
&F_{mml} = 30\pi \tilde{\lambda }^{2}\cot \theta _{\mathrm{o}}\csc^{3} \theta _{\mathrm{o}} -\frac{15}{2} \pi \cot \theta _{\mathrm{o}}\csc \theta _{\mathrm{o}} 
+16\big( -16\tilde{\lambda }^{4}\csc^{6} \theta _{\mathrm{o}} \\
&\ \ \ \ \ \ \ \ \ +4\tilde{\lambda }^{2}( 7\tilde{\lambda }^{2}
+4\tilde{\eta })\csc^{4} \theta _{\mathrm{o}}
-( 13\tilde{\lambda }^{2} +2\tilde{\eta })\csc^{2} \theta _{\mathrm{o}} +1\big) \mu ^{-1},\\
&F_{maa} = -4\tilde{\lambda }\big( (\tilde{\eta } -3\tilde{\lambda }^{2})\csc^{2} \theta _{\mathrm{o}} +2\big),\\
&F_{mal} = 16\tilde{\lambda }\cot \theta _{\mathrm{o}}\csc \theta _{\mathrm{o}}( 3-8\tilde{\lambda }^{2}\csc^{2} \theta _{\mathrm{o}}),\\
&F_{mQQ} = - 6\pi \tilde{\lambda }\cot \theta _{\mathrm{o}}\csc^{3} \theta _{\mathrm{o}} \mu -16\tilde{\lambda }\csc^{2} \theta _{\mathrm{o}},\\
&F_{mll} = 8 \tilde{\lambda }\csc^{6} \theta _{\mathrm{o}}( 12\cos 2\theta _{\mathrm{o}}\tilde{\lambda }^{2} +19\tilde{\lambda }^{2} -\tilde{\eta } +\cos 4\theta _{\mathrm{o}}),\\
&F_{aal} = \big( -4\tilde{\lambda }^{2}(\tilde{\eta } -3\tilde{\lambda }^{2})\csc^{2} \theta _{\mathrm{o}} -( 13\tilde{\lambda }^{2} -3\tilde{\eta } +\cos 2\theta _{\mathrm{o}})\big) \mu ^{-1} ,\\
&F_{aQQ} = \frac{3}{2} \pi \tilde{\lambda }^{2}\csc^{2} \theta _{\mathrm{o}} -\frac{\pi }{2},\\
&F_{all} = 5\pi-15\pi \tilde{\lambda }^{2}\csc^{2} \theta _{\mathrm{o}} -8( 8\cot \theta _{\mathrm{o}}\csc^{3} \theta _{\mathrm{o}}\tilde{\lambda }^{4} -6\cot \theta _{\mathrm{o}}\csc \theta _{\mathrm{o}}\tilde{\lambda }^{2} +\cos \theta _{\mathrm{o}}) \mu ^{-1},\\
&F_{QQl} = \frac{3}{2} \pi \cot \theta _{\mathrm{o}}\csc \theta _{\mathrm{o}} -6\pi \tilde{\lambda }^{2}\cot \theta _{\mathrm{o}}\csc^{3} \theta _{\mathrm{o}}, \\
&F_{lll} = \frac{15}{4} \pi \csc^{2} \theta _{\mathrm{o}}\tilde{\lambda } 
+\frac{15}{2} \pi (\cos \theta _{\mathrm{o}} +\sqrt{\tilde{\eta }})\csc^{2} \theta _{\mathrm{o}}\tilde{\lambda }^{2} \mu ^{-2} \\
&\ \ \ \ \ \ \ +\frac{64}{3}\csc^{2} \theta _{\mathrm{o}}\big( -4\tilde{\lambda }^{2}\csc^{4} \theta _{\mathrm{o}} +3( 2\tilde{\lambda }^{2} +\tilde{\eta })\csc^{2} \theta _{\mathrm{o}}-2\Big)\tilde{\lambda }^{4} \mu ^{-3} \\
&\ \ \ \ \ \ \ +\frac{1}{3}\big( 16\csc^{2} \theta _{\mathrm{o}}
( -12\csc^{2} \theta _{\mathrm{o}}\tilde{\lambda }^{2} +25\tilde{\lambda }^{2} +18\tilde{\eta })\tilde{\lambda }^{4}\\
&\ \ \ \ \ \ \ +( -207\tilde{\lambda }^{4} -90\tilde{\eta }\tilde{\lambda }^{2} +5\tilde{\eta }^{2}) 
-4( 2\tilde{\eta } -\tilde{\lambda }^{2})\cos 2\theta _{\mathrm{o}} +3\cos 4\theta _{\mathrm{o}}\big) \mu ^{-5},\\
&F_{mmQ} = F_{maQ} = F_{mQl} = F_{aaa} = F_{aaQ} = F_{aQl} = F_{QQQ} = 0.
\end{aligned}
\end{equation}

Therefore, we have deduced the expressions of $\Delta\theta$ and $\Delta\phi$ using the geodesic equations. By inserting these expressions back into (\ref{expansion X}) and (\ref{expansion x}), we find that $X-x$ is a third order small quantity with respect to $\tilde{m}$, $\tilde{a}$, $\tilde{Q}$ and $\tilde{l}$. Thus by (\ref{chi approx sine of chi}), $X-x$ is equal to the Faraday rotation angle $\chi$ up to the third order.  

%\nocite{*}

\bibliography{apssamp}% Produces the bibliography via BibTeX.

\end{document}